\documentclass[letterpaper,twocolumn,10pt]{article}
\usepackage{usenix,epsfig,endnotes}
\usepackage{orcidlink}
\usepackage{xurl}

\newcommand{\pykvcache}{\textit{py-kvcache}}
\newcommand{\vllm}{\textit{vLLM}}
\newcommand{\iou}{\textit{io\_uring}}
\newcommand{\lmcache}{\textit{LMCache}}
\newcommand{\llmd}{\textit{llm-d}}
\newcommand{\kvtransfer}{\textit{KV Transfer API}}

\newcommand{\transfer}{\textit{Transfer}}
\newcommand{\kvoffload}{\textit{KV Offload API}}
\newcommand{\kvoffloadname}{\textit{KV Offload}}
\newcommand{\offload}{\textit{Offload}}
\newcommand{\longbench}{\textit{LongBench}}
\newcommand{\scbench}{\textit{SCBench}}
\newcommand{\sharegpt}{\textit{ShareGPT}}
\newcommand{\bailian}{\textit{Bailian}}
\newcommand{\pagedattention}{\textit{PagedAttention}}
\newcommand{\xnvme}{\textit{xNVMe}}
\newcommand{\spdk}{\textit{SPDK}}
\newcommand{\fio}{\textit{fio}}
\newcommand{\liburing}{\textit{liburing}}

\newcommand{\vllmVnine}{\vllmrel{v0.9.0}{v0.9.0}}
\newcommand{\vllmVeleven}{\vllmrel{v0.11.0}{v0.11.0}}
\newcommand{\vllmVsixteen}{\vllmrel{v0.16.0}{v0.16}}
\newcommand{\vllmVeighteen}{\vllmrel{v0.18.0}{v0.18}}
\newcommand{\vllmVnineteen}{\vllmrel{v0.19.0}{v0.19}}

\newcommand{\vllmVtwentytwo}{\vllmrel{v0.22.0}{v0.22}}
\newcommand{\lmcacheV}{\href{https://github.com/LMCache/LMCache/tree/v0.4.7}{v0.4.7}}
\newcommand{\llmdfork}{\href{https://github.com/t348575/llm-d-kv-cache/tree/118c1abe7857d4dab90b4d2be1556ce766700f62}{\#118c1ab}}
\newcommand{\vllmfork}{\href{https://github.com/t348575/vllm/commit/d6eadf416bb5234047760bf55d532f2f038cf697}{\#d6eadf4}}
\newcommand{\pykvcachecommit}{\href{https://github.com/atlarge-research/py-kvcache/tree/790074addbccbe95eda5bc4cfe6afe71d524a910}{\#790074a}}

\begin{document}

\date{}

\title{\Large \bf Building py-kvcache: A Performance Characterization of External KV Caching for vLLM with NVMe SSDs}

\author{
{\rm Joseph Kanichai}\,\orcidlink{0009-0005-7960-7538}\\
Vrije Universiteit Amsterdam\\
Amsterdam, the Netherlands
\and
{\rm Tiziano De Matteis}\,\orcidlink{0000-0002-9158-6849}\\
Vrije Universiteit Amsterdam\\
Amsterdam, the Netherlands
\and
{\rm Animesh Trivedi}\,\orcidlink{0000-0003-3586-7168}\\
IBM Research Zurich\\
Zurich, Switzerland
}

\maketitle

{\renewcommand\thefootnote{}\footnotetext{%
  This work is partially supported by Netherlands-funded projects NWO MLS (OCENW.KLEIN.561).
  This work used the Dutch national e-infrastructure with the support of the SURF Cooperative using grant no.\ EINF-15777.
  This work was carried out between January and July 2026, supervised by Tiziano De Matteis and Animesh Trivedi.
}}

\thispagestyle{empty}

\section*{Abstract}
Prefix caching can reduce the time to first token (TTFT) of long-context LLM requests by reusing previously computed key-value (KV) states, but for short prefixes or fast GPUs, recomputation can be faster than loading from an external cache. We characterize this tradeoff in \vllm{} across GPU, CPU, and NVMe tiers using synthetic workloads, long-context benchmarks, production traces, and find that cache performance depends on transfer granularity, intermediate memory use, and when transfers enter the request schedule, not only on device bandwidth.

These findings motivate \pykvcache{}, a \vllm{} \kvoffloadname{} connector with asynchronous direct I/O, bounded shared staging, and scheduler-aware preloading, which starts disk reads while requests are still waiting, overlapping with compute. At 80k tokens, \pykvcache{} loading from disk is 2.0$\times$ faster than \lmcache{}, with preloading contributing 1.34$\times$. With GPU, CPU, and disk caching enabled, it is 1.23$\times$ faster than \lmcache{} and within approximately 4\% of the native \vllm{} \kvoffloadname{} implementation.

\longbench{} and \scbench{} show that these benefits extend to irregular prefix chains and multi-turn workloads. \bailian{} trace replays improve TTFT on a weaker GPU, but on an H100 the average request falls below the break-even point and GPU memory alone retains enough prefixes. External KV caching should therefore be treated as a setup specific admission decision.

The \pykvcache{} implementation is available at \href{https://github.com/atlarge-research/py-kvcache}{github.com/atlarge-research/py-kvcache}.

\section{Introduction}

Modern large language model (LLM) serving is increasingly a systems problem as much as a model problem. Transformer attention becomes increasingly expensive as context length grows, as during prefill, each token pays ``attention'' to all earlier tokens, making attention computation scale quadratically with prompt length. To avoid recomputing attention for previously processed tokens, each layer's key and value tensors can be stored in a key-value (KV) cache \cite{vaswani2017attention}. This cache can be essential for efficient decoding, but it grows linearly with context length, batch size, model width, and number of layers. As a result, long-context workloads can become limited by both quadratic prefill computation and KV cache capacity and data transfer rates. Recent work on \pagedattention{}, KV cache quantization, and KV cache offloading all treat KV memory as a serious bottleneck in LLM inference systems \cite{kwon2023efficient,liu2024kivi,zheng2026offloadingbottlenecks}.

Prefix caching is a particularly attractive optimization because many real serving workloads contain repeated context. Examples include querying the same long document, multi-turn chat with a system prompt, retrieval-augmented generation (RAG) over repeated passages, and batch evaluation where many questions share a common prefix. In these cases, an inference engine can cache the KV blocks produced during prefill and reuse them for later requests, reducing time-to-first-token (TTFT) as well as the total request time. \vllm{} is an open source LLM serving engine. Its core abstraction, \pagedattention{}, manages the KV cache in blocks, allowing requests with variable length prompts and generations to share GPU memory efficiently while being scheduled together \cite{kwon2023efficient}. \vllm{} exposes repeated prefix reuse through automatic prefix caching and manages KV blocks on the GPU using \pagedattention{} \cite{kwon2023efficient,vllmPrefixCaching}. However, GPU memory is finite, and once the reusable working set exceeds video RAM (VRAM) capacity, the system must either evict useful KV blocks or move KV data to slower but larger tiers.

External KV caching extends prefix caching beyond GPU memory by storing KV blocks in CPU DRAM, local NVMe storage, shared filesystems, or even remote KV stores such as Redis. Systems such as \lmcache{}, \vllm{}'s own external caching mechanisms, and \llmd{} offloading components make the KV cache a distributed multi-tier storage platform rather than a simpler GPU-only data structure \cite{lmcache2025efficient,lmcacheArchitecture,vllmKVOffloading,llmdFSKVCache}. This design is appealing because modern NVMe SSDs offer far better cost per gigabyte when compared to GPU VRAM or traditional DRAM. External caching is not automatically beneficial, since a cache hit replaces GPU prefill computation with cache lookup, disk or host memory reads, and a CPU~$\leftrightarrow$~GPU transfer, not to mention scheduling, synchronization, etc. For short prompts or prompts with low cache hit rates, or when using fast GPUs, these costs can exceed the recomputation time that caching is meant to avoid. This creates a gap in existing systems: they provide mechanisms for storing and retrieving KV blocks from larger tiers, but pair them with a fixed policy that treats every prefix hit as a win. What is missing is a policy that decides when a lookup pays for itself, and writing one requires knowing where the dominant overheads arise and how scheduling decisions interact with I/O and transfer placement.

This work studies that tradeoff in \vllm{}, in various tiers and external KV caching systems. Our starting point is an implementation and measurement effort around the \vllm{} \kvtransfer{}, \lmcache{}, and the newer \vllm{} \kvoffload{}. A significant part of the work is code path tracing, where we followed KV blocks as they move through \vllm{} scheduling logic, KV connector and transfer interfaces, CPU memory, GPU memory, and filesystem/NVMe storage to determine which operations lie on the TTFT critical path. The early experiments showed that API structure matters: blocking transfer paths, the granularity of KV chunks, and whether copies overlap with model forward passes can substantially change TTFT. These are not fixed properties that can be tuned once, since the same configuration wins or loses depending on prompt length, hardware, model, and how long a request waits before execution. Later experiments therefore shifted the focus from raw I/O bandwidth to request scheduling and critical-path placement. In particular, our notes show that reading KV data only after a request is scheduled leaves disk I/O on the TTFT critical path, while preloading data for waiting requests can leave only the final CPU-to-GPU transfer when the request begins service.

Based on these observations, we built \pykvcache{}, an external Python KV-cache backend designed for \vllm{} and filesystem/NVMe storage. The design emphasizes four principles. First, the cache should be simple to deploy and share across multiple \vllm{} instances by using an ordinary filesystem layout. Second, the I/O path should be efficient for the large, contiguous KV objects produced by block-based serving engines, avoiding excessive worker-thread coordination when a smaller number of outstanding large reads and writes is sufficient. Third, intermediate CPU memory should be explicitly bounded rather than growing with request size. Fourth, cache policy should be hardware and workload aware: external KV reuse should be bypassed below the measured break-even point instead of treating every prefix hit as a win.

The main question we ask is therefore not merely whether external KV caching can work, but \textit{when it should be used}. Our experiments compare \lmcache{}, \vllm{} \kvoffloadname{}, and \pykvcache{} across synthetic long-document workloads and trace-derived workloads. The results preview a nuanced answer. For long repeated documents, \pykvcache{} with preload reduces TTFT substantially relative to \lmcache{} and to \pykvcache{} without preload. In the tiered configuration at 80k-token document sizes, \pykvcache{} is 1.23$\times$ faster than \lmcache{} and stays within 1.04$\times$ of the native \vllm{} \kvoffloadname{} implementation. Preload is essential: in the disk-only configuration it halves TTFT relative to \lmcache{} at 80k tokens, and against \pykvcache{} without preload it lowers TTFT by 1.34$\times$ at 80k tokens and by 1.66$\times$ at 40k tokens. However, \bailian{} trace experiments indicate that some real workloads have average request sizes below the break-even point for our setup, so external KV caching offers no TTFT benefit over GPU prefix caching alone despite cache hits.

This work makes the following contributions:

\begin{itemize}
    \item We trace and characterize the performance behaviour of \vllm{}'s \kvtransfer{} and \kvoffload{} and related KV-cache systems, including how scheduling, copy granularity, and copy/compute overlap affect TTFT.
    \item We design \pykvcache{}, a Python external KV cache engine for \vllm{} for use with a shared filesystem, utilizing \iou{}.
    \item We introduce and evaluate a preload mechanism that reads KV data for waiting requests before they enter execution, shifting disk I/O off the request critical path, and improving overlap of copy and compute.
    \item We show that external KV caching has hardware and workload dependent break-even points, motivating cache admission and bypass policies instead of unconditional loading and storing.
\end{itemize}

\section{Background}

This section provides the background needed to understand external KV caching, and how they work in \vllm{}. We first review how transformer inference uses the KV cache during prefill and decoding, why the cache becomes a major memory object for long-context workloads, and how prefix reuse changes the cost of repeated requests. We then describe why serving systems move KV data beyond GPU memory and discuss the \vllm{} interfaces that allow this.

\subsection{KV Cache in LLM Serving}

Autoregressive transformer models generate sequences of data, essentially predicting the next element based on all previously generated elements. Autoregressive transformer inference has two phases: \textit{prefill} and \textit{decode}. During prefill, the model processes the input prompt and computes a vector representation for each token at each layer, with each vector formed by attending to the tokens that came before it. For a prompt of length $L$, because each token is compared against all previous tokens, this work grows quadratically with $L$ \cite{vaswani2017attention}. During decode, the model generates one new token at a time based on the previous prefill. Without additional state, each decode step would need to recompute attention over the entire prefix, making generation expensive for long prompts.

The KV cache avoids this recomputation. In each transformer layer, the attention mechanism projects hidden states into queries, keys, and values. Once a token has been processed, its key and value tensors can be stored and reused by later tokens. This reuse applies only to tokens in the prefix, since newly generated tokens depend on the model's previous outputs, so their KV tensors do not exist until generation reaches them. This also means that even though two prompts share identical text in the middle of a prompt, they cannot be interchanged unless the preceding text is also identical. A decode step then computes the query for the new token and attends over the cached keys and values from the prefix, rather than recomputing those keys and values from the original tokens.

The KV cache grows linearly with the number of cached tokens and is calculated like so: $\mbox{bytes per token} = \mbox{layers} \times 2 \times \mbox{KV heads} \times \mbox{head dim} \times \mbox{datatype size}$. As such, even a small model like Llama~3.2~3B, the bytes per token is 112~KiB, so a context of 80k tokens takes up 8.75~GiB. This can quickly overwhelm the limited GPU VRAM. It also creates a data movement problem since loading, storing, evicting, transferring KV tensors can directly affect request latency if those operations sit on the critical hot path for inference.

Most LLM serving systems manage the KV cache by placing groups of tokens in blocks rather than as one tensor per request. Block management makes it easier to allocate memory for variable length prompts and generations and evict or transfer cache regions at a useful granularity. \vllm{}'s \pagedattention{} is a popular implementation of this approach, using fixed size KV blocks to reduce fragmentation and support high throughput batched serving \cite{kwon2023efficient}. This block abstraction is the basis for the external cache systems studied in this work.

To identify reusable KV blocks, \vllm{} computes a hash for each full token block. The hash uses both the tokens in the current block and the hash of the preceding block, forming a chain from the start of the prompt. Two requests can therefore reuse a block only when their token sequences match from the beginning of the prompt through that block. Identical text appearing later in otherwise different prompts does not produce a reusable match. Cache lookup proceeds block by block until the first mismatch, and only the contiguous matching prefix is loaded. This makes prefix reuse and cache granularity dependent on the configured block size, since a partially matching block must still be recomputed.

\begin{figure}[t]
  \centering
  \includegraphics[width=\columnwidth]{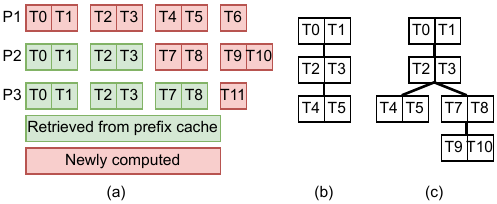}
  \caption{Prefix-hash chaining.}
  \label{fig:prefix-hash-example}
\end{figure}

\autoref{fig:prefix-hash-example} illustrates this process with two-token blocks. In panel (a), P1 has no reusable prefix and computes every token. Its three full blocks form the hash chain in panel (b), while the partial block containing T6 is not cached. P2 then reuses the first two blocks before branching to T7--T8 and T9--T10, producing the prefix tree in panel (c). P3 can follow that second branch through T7--T8 and needs to compute only its unmatched partial block, T11. Blocks before a branch can therefore share KV data, whereas blocks following different token sequences receive different chained hashes.

\subsection{Extending KV Cache Beyond GPU Memory}

GPU memory is the best place to keep KV blocks, since it offers the highest bandwidth, but it is also the smallest and most expensive tier. Once the active KV working set exceeds GPU memory, the serving engine must either evict blocks and recompute them later or move the KV cache to a larger but slower storage tier. External KV caching does just this, by storing on other media outside the GPU and loading back when a later request reuses the same prefix.

The closest external tier is CPU DRAM. DRAM has much larger capacity than GPU memory and can be shared by the host process, but using it still requires CPU~$\leftrightarrow$~GPU transfers over PCIe. Because of the general design of all modern computer hardware, it is not possible to escape the PCIe link, and all secondary tiers for a KV cache must pass through the PCIe bus (excluding tightly coupled chips such as Nvidia's Grace Hopper platform). Each secondary tier used for storing KV cache adds copy, metadata, synchronization costs.

Using storage as a secondary tier provides far greater capacity thanks to the far lower cost per gigabyte offered by modern NVMe SSDs when compared to DRAM or GPU VRAM. Shared or distributed filesystems take this idea further allowing whole clusters and deployments to share their secondary cache, which should result in better cache hit rates.

\lmcache{} adds external KV cache storage for \vllm{} and supports moving KV data across GPU, CPU, local storage, and remote or shared backends such as object stores like S3, or Redis \cite{lmcache2025efficient,lmcacheArchitecture}. \vllm{}'s \kvoffload{} provides a native interface for offloading KV blocks from GPU memory and overlapping transfer work with model execution \cite{vllmKVOffloading}, and offers its own DRAM and filesystem tier. Numerous other external KV cache stores exist such as \llmd{} \cite{llmdFSKVCache} and Mooncake \cite{qin2025mooncake}. These systems make external KV caching practical, but they also motivate characterization of external KV caching, to understand tradeoffs and how they react to changes in setup and workload.

\subsection{vLLM KV Cache Interfaces}

\vllm{} exposes external KV caching through interfaces that let a connector load KV blocks for a request and store newly produced KV blocks for later reuse. The lower-level interface, called the \kvtransfer{}, separates cache management into scheduler and worker operations. The scheduler methods run in the \vllm{} scheduler process and operate on request metadata rather than tensor data, by computing token or block hashes, checking whether a prefix is present in the external cache, and deciding how many prefix tokens can be loaded. The worker methods run in the GPU worker process and perform the actual data movement, both loading matched KV tensors into GPU cache slots and storing newly generated KV tensors into the external tier. This split is at the core of how \vllm{} is designed, since the scheduler does not run GPU code, or have control over the GPU tensors.

\begin{figure*}[p]
  \centering
  \begin{subfigure}[t]{0.58\textwidth}
    \centering
    \includegraphics[width=\linewidth]{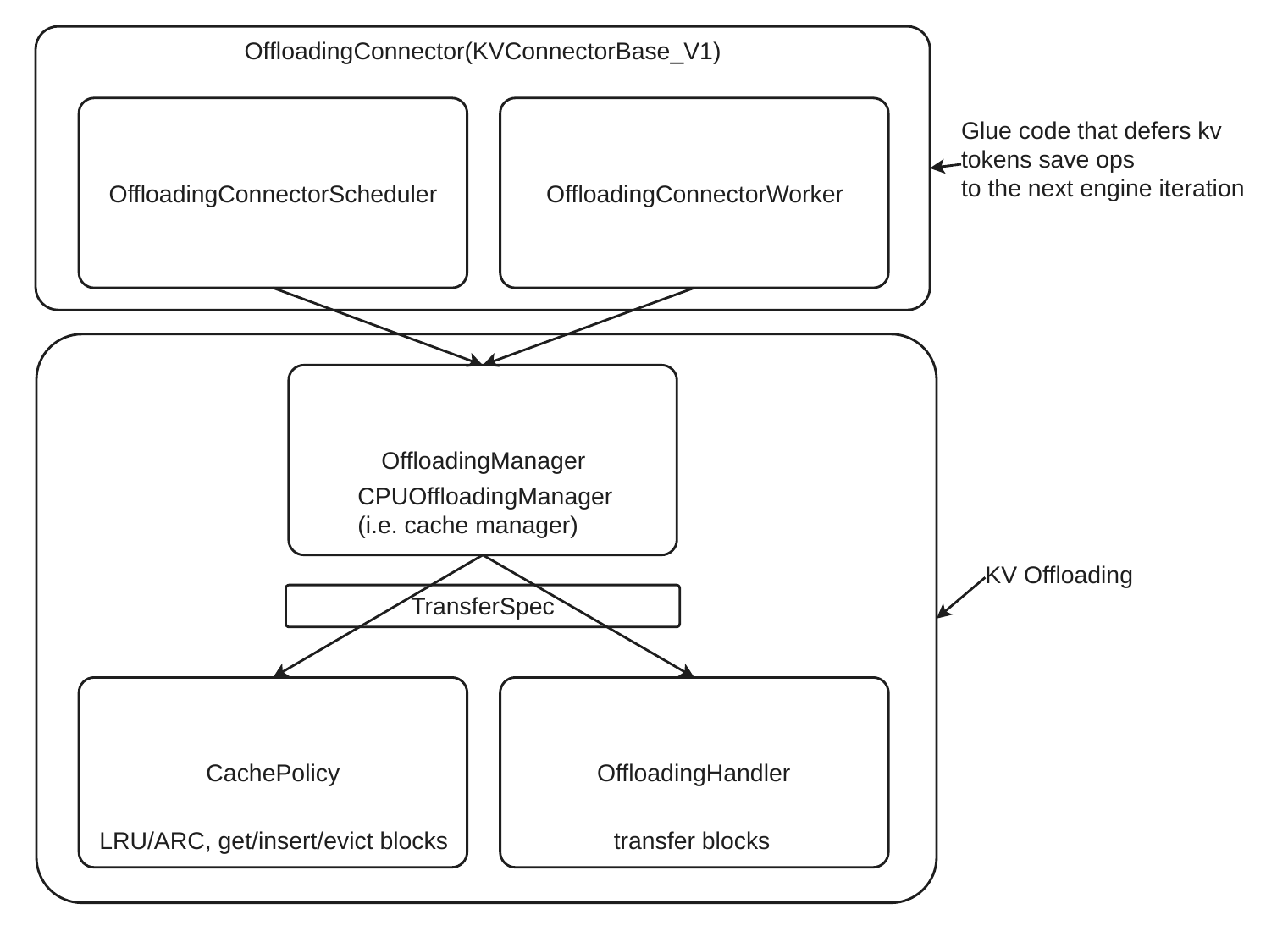}
    \caption{\kvoffload{} wrapper classes.}
    \label{fig:kv_offload_classes}
  \end{subfigure}

  \smallskip
  \begin{subfigure}[t]{0.72\textwidth}
    \centering
    \includegraphics[width=\linewidth]{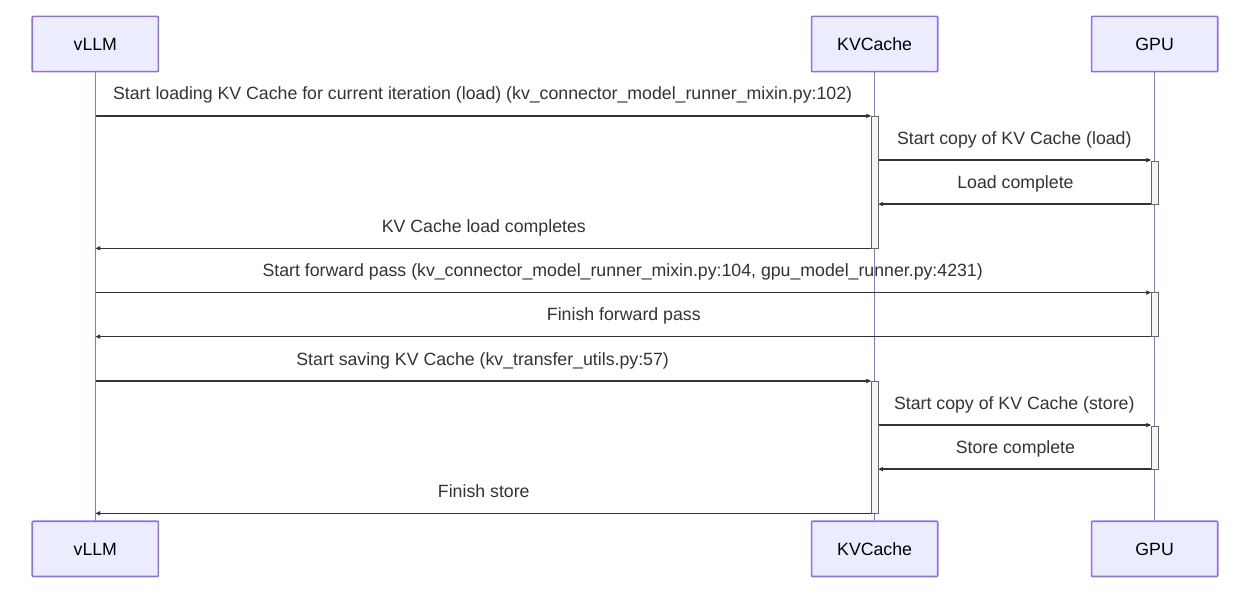}
    \caption{Synchronous \kvtransfer{} execution sequence.}
    \label{fig:kv_transfer_detailed}
  \end{subfigure}

  \smallskip
  \begin{subfigure}[t]{0.72\textwidth}
    \centering
    \includegraphics[width=\linewidth]{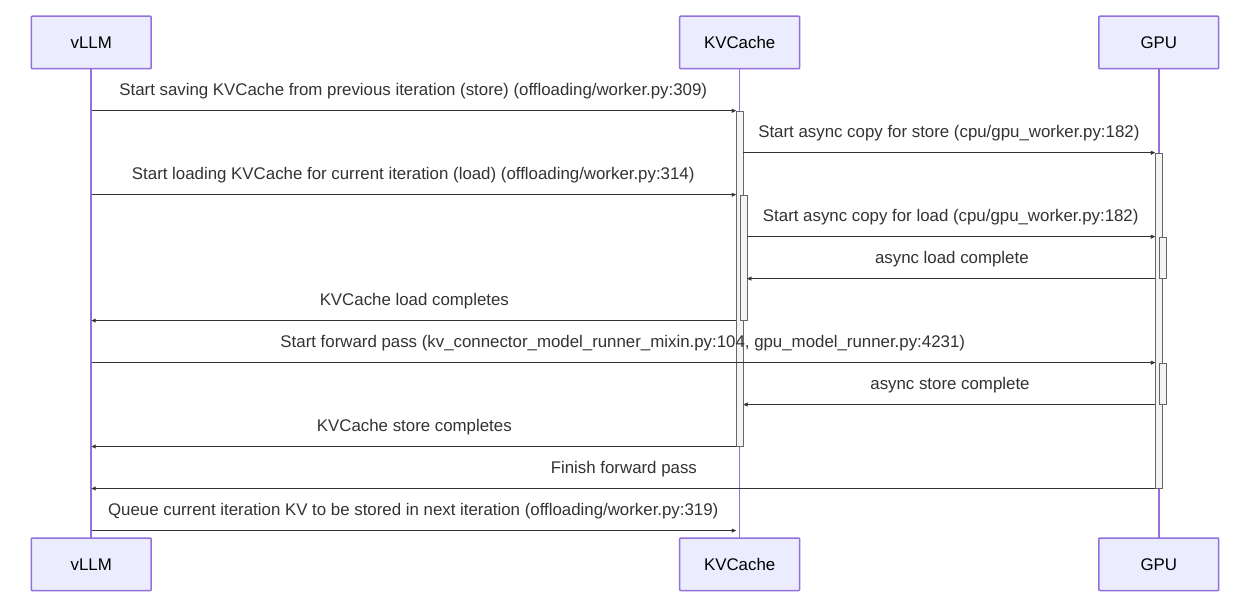}
    \caption{\kvoffload{} execution sequence.}
    \label{fig:kv_offload_detailed}
  \end{subfigure}
  \caption{\vllm{} KV cache interfaces and execution sequences (v0.22.0). \offload{} defers stores to the next engine iteration so they overlap the following forward pass, while \transfer{} waits for them after each pass.}
  \label{fig:api_flow_comparison}
\end{figure*}

The \transfer{} connector was synchronous until \vllmVnine{}\footnote{\url{https://github.com/vllm-project/vllm}}, so \vllm{} could block while a connector loaded or stored KV data. v0.9.0 added asynchronous support, but the API still requires the connector to manage request completion. In practice, this makes the \kvtransfer{} complex to implement, and requires intimate knowledge of data flow and system state, resulting in numerous methods that need to be implemented, all with minimal documentation. Most existing external KV cache systems like \lmcache{} use the \transfer{} connector as shown in \autoref{fig:kv_transfer_detailed}.

The \kvoffload{} introduced in \vllmVeleven{} provides a smaller abstraction for the same general problem. Instead of requiring a connector to implement many scheduler and worker hooks, \offload{} exposes a few simpler operations: \texttt{prepare\_load} and \texttt{prepare\_store} on the scheduler side, and \texttt{transfer\_async} on the worker side. Internally, \vllm{} implements this API as a wrapper over the \kvtransfer{}. An \textit{offloading\_connector} translates the simpler offload operations into the lower-level \transfer{} interface, as shown in \autoref{fig:kv_offload_classes}. \offload{} allows implementors to only think about data movement and cache management, without worrying about async state management, request lifecycle and other similar actions required when implementing the \kvtransfer{}. The important systems difference is that \offload{} is designed around asynchronous movement of KV blocks. In particular, it defers store work from engine iteration $N$ to iteration $N+1$, so GPU~$\rightarrow$~CPU copies for the previous iteration can overlap with the model's forward pass for the next iteration. \autoref{fig:kv_offload_detailed} shows the execution sequence of \offload{}.

Unlike the \kvtransfer{}, \vllm{} provides its own implementation of connectors using \offload{}, by providing a DRAM tier, with a cache manager running either least recently used (LRU) or adaptive replacement cache (ARC) eviction for management. \vllm{} have also introduced ``Secondary Tiers'', one of which offloads to a filesystem, described further in \autoref{subsec:native-offload}.

\subsection{Metrics}

The main latency metric in this work is time-to-first-token (TTFT), the time from when a request is submitted until the first output token is produced. TTFT is especially important for interactive workloads because it includes queueing, scheduling, prefill or cache loading, GPU transfer, and the first decode step. External KV caching primarily affects TTFT because a cache hit replaces some or all of prefill with cache lookup and KV movement. In benchmarks where multiple requests are sent at the same time, other requests can spend time waiting, changing the TTFT.

A system with high storage bandwidth can still perform poorly if cache reads are placed late in the request path, and a slower tier can be useful if its work is started early enough. We therefore report not only end-to-end TTFT, but where applicable we also analyze total benchmark runtime, cache lookup time, CPU~$\leftrightarrow$~GPU transfer times, disk read/write time and model execution. We also analyze bandwidth of these operations in GB/s. These measurements let us identify whether an optimization reduces total work, moves work off the critical path, or merely shifts overhead from one component to another.

For cache behaviour, we report cache hit rate and break-even point. Cache hit rate measures what fraction of requests, tokens, or KV blocks can be served from an existing cache entry rather than recomputed. We also discuss a break-even point, which captures the minimum prefix size where loading cache KV data is faster than recomputing the same prefix on the GPU.

\begin{table*}[ht]
  \centering
  \small
  \setlength{\tabcolsep}{4pt}
  \begin{tabular}{|p{0.11\textwidth}|p{0.19\textwidth}|p{0.19\textwidth}|p{0.19\textwidth}|p{0.19\textwidth}|}
    \hline
     & \lmcache{} & \llmd{} & Native \vllm{} \offload{} & \pykvcache{} \\
    \hline
    Version & \lmcacheV{} & Our direct I/O fork, based on v0.8 \llmdfork{} & Our instrumented fork \vllmfork{} & \pykvcachecommit{} \\
    \hline
    Tested with \vllm{} version & \vllmVsixteen{}, \vllmVtwentytwo{} & \vllmVeighteen{}, \vllmVnineteen{}, \vllmVtwentytwo{} & \vllmVsixteen{}, \vllmVtwentytwo{} & \vllmVtwentytwo{} \\
    \hline
    \vllm{} KV Cache  API & \kvtransfer{} & \kvoffload{} & \kvoffload{} & \kvoffload{} \\
    \hline
    Supported KV Cache tiers & CPU DRAM, disk, remote object stores, P2P & CPU DRAM (for staging), disk & CPU DRAM, disk (remote object stores and P2P only after v0.22) & CPU DRAM, disk \\
    \hline
    Disk I/O & POSIX, 4-thread pool & \texttt{O\_DIRECT} in our fork, thread pool with read or write preferring workers & \texttt{O\_DIRECT}, thread pool with read or write preferring workers & \iou{}, single thread with bounded I/O depth \\
    \hline
    Staging memory & CPU tier's pinned pool doubles as staging & One pinned buffer per I/O thread & CPU tier doubles as staging & Shared with CPU tier \\
    \hline
    Filesystem layout & Flat directory, hashed filenames & Hierarchical, hash-addressed & Same scheme as \llmd{} & Same scheme as \llmd{} \\
    \hline
    Language & Python & C++ & Python & Python \\
    \hline
  \end{tabular}
  \caption{Summary of the compared external KV cache systems. \pykvcache{} is this work, and is described in \autoref{sec:design}.}
  \label{tab:systems-compared}
\end{table*}

\section{Systems Compared}

We compare \pykvcache{}, our own external KV cache, against three existing systems. \autoref{tab:systems-compared} summarizes their versions, \vllm{} interfaces, and storage designs, and the subsections below describe three existing systems, while \pykvcache{} is described in \autoref{sec:design}.

\subsection{LMCache}
For our experiments, we identified \lmcache{} (\lmcacheV{}) as the primary external KV cache with a disk tier \cite{lmcache2025efficient,lmcacheArchitecture}. \lmcache{} is a popular external KV Cache, supporting multiple store types, GPUs, inference engines besides \vllm{}, distributed P2P deployments, integration with Kubernetes and much more. It supports being run as only a CPU DRAM KV Cache, and also with a secondary slower tier such as with disk or remote stores. For the disk tier, we used \lmcache{}'s default disk tier (although they have recently introduced a native disk tier as well). The disk tier uses a small thread pool and standard POSIX file operations to write all cache files to the same directory, by hashing them with the parameters of the model and the prefix hash.

\subsection{llm-d}
\label{subsec:llm-d}
For comparative purposes, we also picked \llmd{}, which provides a filesystem KV Cache \cite{llmdFSKVCache}.\footnote{\url{https://github.com/llm-d/llm-d-kv-cache}} However, this was deprecated while our experiments were ongoing, in favour of the secondary filesystem disk tier introduced directly into \vllm{}. That tier is functionally identical to what \llmd{} provided, keeping the same on-disk format and the same direct I/O thread-pool design, so it replaces \llmd{} rather than changing what is being measured. As a result, some of our Pareto fronts were conducted in \llmd{}. \llmd{}-kv-cache is a KV cache written in C++, that uses a staging CPU memory pool as an intermediate point for FS operations. It used a thread pool, where workers were reserved for read or write operations. \llmd{} uses standard POSIX file read and write operations, going through the page cache. However, for our tests this is unfavourable, and is likely to distort results, especially when cached data is re-read. To prevent this, we built a fork of \llmd{} with direct I/O support (based on v0.8.0-fs-v0.20, \llmdfork{})\footnote{\url{https://github.com/t348575/llm-d-kv-cache}}. \llmd{} also maintains a hierarchy of folders and subfolders based on model parameters and the prefix hash, with the aim of reducing filesystem contention by limiting the number of files in each directory.

\subsection{Native vLLM Offload Implementation}
\label{subsec:native-offload}
\vllm{}'s \kvoffload{} includes a native Python implementation with a CPU DRAM cache managed by an ARC or LRU eviction policy, and optional secondary tiers for filesystem, object stores or P2P transfer \cite{vllmKVOffloading}. Multi-tier offloading was introduced in v0.22.0, and is orchestrated by the \texttt{TieringOffloadingManager}. The CPU tier acts as a gateway between secondary tiers and the GPU. On a store, \vllm{} first copies a completed GPU KV block to the CPU primary tier and then asynchronously propagates it to the secondary tiers. The CPU block remains protected from eviction until the filesystem write completes. On a lookup, \vllm{} checks the CPU tier first. A secondary tier hit is promoted into a reserved CPU slot, after which it can be copied to the GPU. While this promotion is in progress, the scheduler defers the request and retries in a later scheduling iteration rather than blocking on the operation.

The filesystem tier uses the same storage format as \llmd{}. It features two thread pools, giving reads and writes separate preferred priority, while allowing idle workers to process the other queue. This is functionally identical to \llmd{}'s staged memory and filesystem design, although the tier manager and cache policy are integrated into \vllm{}.

\section{Experimental Setup}
\label{sec:experimental-setup}

This section describes how we evaluate external KV caching systems and why each experiment was chosen. Our goal is not only to compare TTFT, but also to break down where the latency comes from. We therefore combine benchmarks with code tracing, to dive deeper into the causes. This work was carried out alongside upstream \vllm{} development, so the experiments span several \vllm{} versions rather than a single release. The evaluation of \pykvcache{} (and comparisons with Native \vllm{} \kvoffload{}, \lmcache{}) in \autoref{sec:evaluation} uses our \vllm{} fork based on \vllmVtwentytwo{}. Most of the characterization in \autoref{sec:characterization} predates that fork. The interface comparison in \autoref{fig:longdoc_ttft} and \autoref{fig:from_gpu_bw} was measured on \vllmVsixteen{}, the \llmd{} connector tracing on \vllmVeighteen{}, and the break-even frontiers in \autoref{fig:other-kv-pareto} and \autoref{fig:min-bandwidth} on \vllmVtwentytwo{}. \autoref{tab:systems-compared} lists the versions used for each system. Experiments use Llama~3.2~3B Instruct and Qwen3~4B Instruct~2507 \cite{llama32_3b_instruct,qwen3_4b_instruct}.

\vllm{} changed substantially over the course of this work, and the version spread follows that development rather than a choice on our part. Our experiments span \vllmVsixteen{} through \vllmVtwentytwo{}. The \kvoffload{} was still new when we began, and was later rewritten to support multiple tiers, so we moved our fork forward as those changes landed. The same development also removed the need for \llmd{}. Its filesystem KV cache was deprecated in favour of the secondary filesystem tier built directly into \vllm{}, which is functionally identical to it, reusing the same on-disk format and disk I/O design. \llmd{} therefore appears in the characterization, where it was the filesystem cache available at the time, while the native \vllm{} filesystem tier takes its place as the architectural comparison point in \autoref{sec:evaluation}. Because the two occupy the same role and the connector has since been deprecated, we stopped testing \llmd{} rather than carrying it through the later experiments.

The experiments are designed to separate three questions:
\begin{itemize}
    \item When cached KV reuse is faster than recomputation
    \item How different \vllm{} cache interfaces perform
    \item Where current KV caches leave performance unexploited
\end{itemize}

\subsection{Benchmarks and Workloads}

For evaluating external KV caches, we used both synthetic and trace driven workloads, representing the current state-of-the-art for testing LLM serving systems as well as workloads for specifically testing and benchmarking the performance of KV caches. These workloads were chosen based on their popularity, as well as their prominence in literature. The synthetic workloads give us maximum configurability, allowing us to control prefix length (how much of a prompt prefix is reused across requests), reuse rate (how many requests use the prefix), concurrency, among numerous other parameters. These synthetic workloads allow understanding the performance of KV caches under highly specific scenarios. Trace driven workloads on the other hand are meant to test whether the same conclusions and observations from running synthetic benchmarks hold under more realistic request patterns.

\subsubsection*{Measurement protocol}
Document and prompt sizes are quoted in binary units throughout this work, so 1k denotes 1{,}024 tokens and 80k denotes 81{,}920 tokens. Each benchmark configuration was repeated three times, and we report the mean across repetitions unless otherwise stated. All tests were run with a single output token unless stated otherwise. Output tokens are generated by the LLM after prefill. We have set output tokens to 1 since any greater number would involve more GPU forward passes. This ensures that our performance results focus on the KV cache and not on long decode steps. All experiments were also run with FP16 KV data, to test the worst case scenario for KV cache size.

\subsubsection*{Long-document}
The long-document benchmark is our main synthetic workload. Versions are included in the \href{https://github.com/vllm-project/vllm/blob/d6eadf416bb5234047760bf55d532f2f038cf697/benchmarks/benchmark_long_document_qa_throughput.py}{\vllm{} benchmark suite}\footnote{\url{https://github.com/vllm-project/vllm/blob/d6eadf416bb5234047760bf55d532f2f038cf697/benchmarks/benchmark_long_document_qa_throughput.py}} and the \href{https://github.com/LMCache/LMCache/blob/v0.4.7/benchmarks/long_doc_qa/long_doc_qa.py}{\lmcache{} benchmark suite}\footnote{\url{https://github.com/LMCache/LMCache/blob/v0.4.7/benchmarks/long_doc_qa/long_doc_qa.py}} and are largely identical. We created our own version with additional control parameters. Some of the parameters supported by this benchmark are: document size, prefix reuse percent, prefix size, concurrency, arrival rate, warmup requests. Each run has a pre-warmup phase that sends a few small requests to \vllm{}. This is to warm up the code path, as well as to soak up any initialization costs. In our tests, we noticed a 2$\times$--5$\times$ higher TTFT for the first request sent to \vllm{} after startup. After this, a warmup phase populates the cache and a query phase sends the requests that reuse it. In our experiments we sweep through a number of prefix reuse rates and document lengths. Our benchmark also supports providing a range of values for almost all parameters.

To evaluate scheduling and overlap, we use mixed workloads with both fresh requests and prefix reuse requests. These runs vary request concurrency, reuse fraction, and the gap between warmup and query traffic. They are used to test whether KV caches can move disk I/O out of the TTFT critical path and whether the benefit remains when the system is serving other requests at the same time.

\subsubsection*{Pareto capture}
\label{sec:pareto}
Because SSDs are the slowest form of memory in our external KV cache chain, it is important to understand when the cost of storing KV on an SSD is worthwhile. To do this, we have created a program that runs experiments using the long-document benchmark on a range of document sizes with KV cache disabled to understand pure GPU computation cost. Next, we re-run the same experiments while using our KV cache. We can do the same for a pure CPU KV cache as well. With the experiment data, we can then interpolate TTFT times for values not explicitly measured (i.e. in-between values).

We define the terms used by the Pareto capture tool as follows:
\[
\begin{array}{rcl}
D & = & \mbox{total document size in tokens}, \\
P & = & \mbox{cached prefix size in tokens}, \\
f(x) & = & \mbox{cold TTFT (full compute)}, \\
g(x) & = & \mbox{cache hit TTFT}.
\end{array}
\]

Caching is worthwhile when:
\[
g(P) + f(D - P) < f(D)
\]

Calculating the Pareto frontier is essential, since it allows us to not only calculate a break-even Pareto front, but also allows us to calculate the minimum break-even data transfer bandwidth for prefix caching to be beneficial.

\subsubsection*{Filesystem metadata}
\label{sec:fs_metadata}
External KV caches that use filesystems do not only depend on sequential read and write bandwidth. They also perform metadata operations such as checking whether a cache object exists, opening files, publishing newly written cache entries, evicting old entries. To isolate these costs, we built a minimal filesystem metadata microbenchmark that measures throughput and latency of various filesystem operations under various configurable parameters.

Each run prepares a populated workspace, performs an unmeasured warmup, and then runs the selected operation for a fixed duration across multiple worker threads. Payload-bearing operations use \texttt{O\_DIRECT} to bypass buffered file-data caching, with optional file and directory \texttt{fsync()}. Some of the operations performed include \texttt{stat}, \texttt{access}, \texttt{open}/\texttt{close}, \texttt{create}/\texttt{unlink}, \texttt{rename}, \texttt{readdir}. These are measured both individually and as the transactions a cache connector actually performs, categorized as \textit{lookup} or \textit{publication} operations. This microbenchmark was deployed to understand if the standard metadata operations in a filesystem can turn into a bottleneck at scale.

\begin{figure*}[t]
  \centering
  \includegraphics[width=0.92\textwidth]{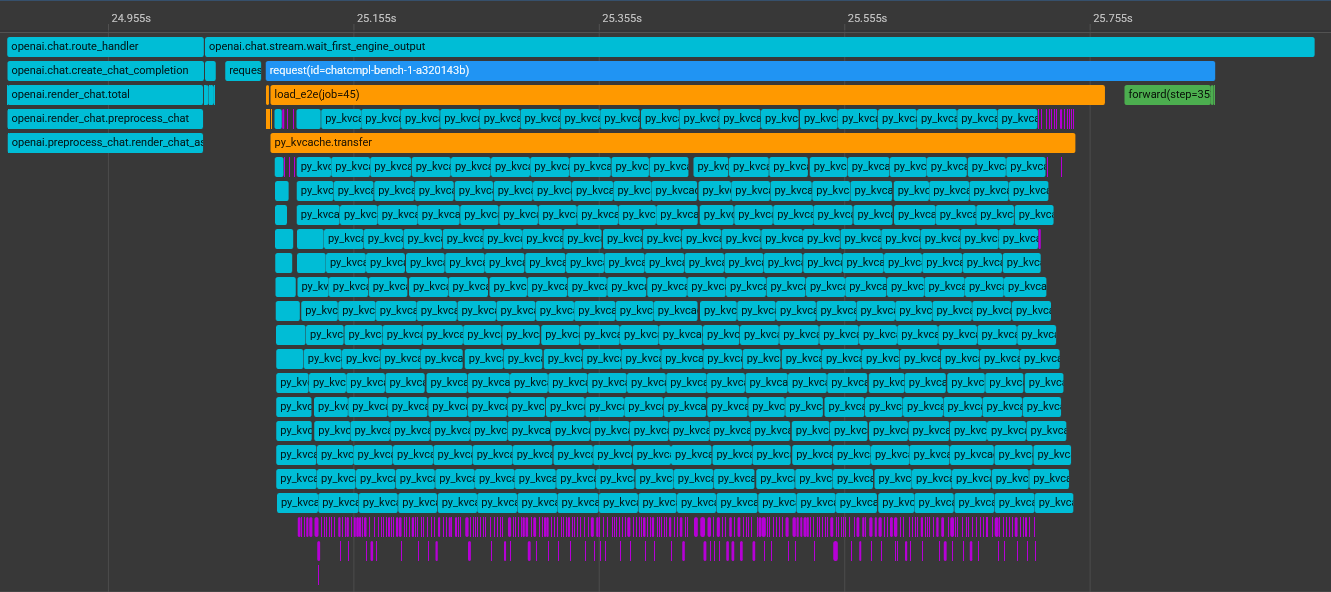}
  \caption{Sample \pykvcache{} execution trace. Blue blocks are I/O reads and pink blocks are transfers to the GPU.}
  \label{fig:trace-sample}
\end{figure*}

\subsubsection*{ShareGPT}
\sharegpt{} is a widely used collection of anonymized, real world ChatGPT conversations \cite{sharegptDataset}. We use \vllm{}'s \sharegpt{} benchmark using \texttt{vllm bench}. The benchmark samples prompts and target response lengths from the dataset and replays them at a configurable request rate or concurrency. It therefore captures a realistic mix of short and medium requests, rather than a single fixed prompt length as in our synthetic workloads.

Unlike the long-document workload, \sharegpt{} does not intentionally construct a shared prefix across requests, and conversations are sampled independently and reuse arises incidentally. We use it as a general serving baseline and to quantify the overhead of enabling an external KV cache when hits are infrequent. The dataset has a low level of prefix reuse, and consists mostly of small conversations and prompt sizes. This distinction is important when interpreting the results, since \sharegpt{} measures behaviour under representative, mostly cold-cache traffic, whereas the controlled workloads isolate the potential benefit of prefix reuse. As a result, \sharegpt{} is only done to observe how external KV caching behaves in these scenarios, and not to establish any definitive performance baseline. As with our other experiments, output tokens is set to 1.

\subsubsection*{Bailian}
\bailian{} is a set of anonymized traces created from samples of production traffic to a Qwen serving cluster on Alibaba Cloud \bailian{} \cite{bailian,bailianDataset}. Similar to \sharegpt{}, this allows us to test realistic requests in size, prefix reuse and arrival. We replay the Coder, interactive (A), and API-driven (B) traces. Our replay driver reconstructs a synthetic prompt, similar to the \sharegpt{} benchmark.

\subsubsection*{SCBench}
\scbench{} (Shared Context Bench) is a multi-turn benchmark to evaluate long-context scenarios, and is specifically designed for evaluating KV caching performance \cite{scbench,scbenchDataset}. Each session contains a context followed by a number of questions. The questions from each session share a prefix, making it possible to test how KV caches handle realistic multi-turn workloads. We do not directly use \scbench{}, but instead only use its dataset. We replay the dataset with our own driver which supports replaying the messages in round-robin format which allows evaluating a worst case scenario when the dataset is much larger than available GPU VRAM and CPU DRAM.

\subsubsection*{LongBench}
\longbench{} is a long-context benchmark based on multiple-choice questions over long documents \cite{longbenchV2,longbenchV2Dataset}. Unlike \scbench{}, \longbench{} focuses on evaluating LLM output performance and quality. We therefore use \longbench{} only for its dataset and prefix reuse.

\subsubsection*{Benchmark Orchestration \& Other scripts}
We use a Python orchestration script to run all benchmark configurations consistently. The script starts \vllm{} with the selected KV cache configuration, executes the requested workload, and collects aggregate and individual request results. Benchmark and server parameters are defined in JSON configuration files, which can expand into parameter sweeps over specified configurations. This provides a common execution path for all systems.
The benchmark orchestration script, driver code for all the benchmarks, along with all the other scripts and plotting for this work are available at \href{https://github.com/atlarge-research/kvcache-experiments}{github.com/atlarge-research/kvcache-experiments}.

\subsubsection*{Tracing \& Profiling}
\label{subsec:tracing}
To understand how the two APIs work, as well as some of the internals of \vllm{}, we implemented \href{https://github.com/t348575/simple-profiler/}{simple-profiler}\footnote{\url{https://github.com/t348575/simple-profiler}}, a profiling/tracing system in Python and a web viewer for analyzing the recorded traces. The tracing library records the start and end times of functions and events, together with supplementary data. Entire functions can be instrumented with \texttt{@profile}, while individual code blocks use \texttt{@profile\_scope}. Asynchronous events that outlive a function call are timed manually and recorded with \texttt{add\_event}. Traces are saved as JSON when the program exits and can then be loaded into the web viewer. \autoref{fig:trace-sample} shows an example subsection of a \pykvcache{} trace rendered by this viewer. Here \texttt{load\_e2e} covers the cache work for one request, and \texttt{forward} only begins once it completes. The smaller blue blocks indicate I/O read operations, and the pink blocks are transfers to the GPU.

\subsection{Hardware \& Software Configuration}

Our experiments were run across two machines. A local node equipped with less powerful GPUs, and GPU nodes on Snellius, the Dutch national supercomputer \cite{snellius}, used to represent more powerful systems. The local node was used for development, debugging, testing, and benchmarks, while the Snellius nodes were used only for benchmarks. Snellius nodes and our local node were allocated exclusively, so no other job shared the GPUs, the offload SSD, or the PCIe links during a measurement. Although the nodes have multiple GPUs, all our experiments use a single GPU. The two systems differ enough in GPU speed and VRAM capacity to fall on opposite sides of the measured break-even point, which \autoref{sec:evaluation} uses to test when external caching stops being beneficial.

\begin{table}[h]
  \centering
  \small
  \begin{tabular}{p{0.27\columnwidth}p{0.58\columnwidth}}
    \hline
    Component & Configuration details \\
    \hline
    Server & Supermicro SYS-221H-TNR \\
    CPU & Intel(R) Xeon(R) Silver 4514Y \\
    DRAM & 256~GiB, DDR5 \\
    GPUs & 2$\times$ Nvidia RTX 4000 Ada (PCIe~4.0), 20~GiB GDDR6 \\
    Offload SSD & Kioxia CM7-R 1.92~TB (PCIe~5.0), Samsung PM9A3 1.92~TB (PCIe~4.0) \\
    OS \& Kernel & Ubuntu 24.04 with kernel 6.8.0 \\
    Runtime stack & CUDA 13.2, Python 3.12, PyTorch 2.11 \\
    \hline
  \end{tabular}
  \caption{Local node configuration.}
  \label{tab:local_machine_config}
\end{table}

\begin{table}[h]
  \centering
  \small
  \begin{tabular}{p{0.27\columnwidth}p{0.58\columnwidth}}
    \hline
    Component & Configuration details \\
    \hline
    Server & ThinkSystem SD665-N V3 \\
    CPU & Dual socket AMD EPYC 9334 \\
    DRAM & 768~GiB, DDR5 \\
    GPUs & 4$\times$ Nvidia H100 (PCIe~5.0), 94~GiB HBM2e \\
    Offload SSD & Samsung PM1743 7.5~TB (PCIe~5.0) \\
    OS \& Kernel & RHEL 9.6 with kernel 5.14.0 \\
    Runtime stack & CUDA 12.8, Python 3.13, PyTorch 2.11 \\
    \hline
  \end{tabular}
  \caption{Snellius GPU node configuration.}
  \label{tab:snellius_machine_config}
\end{table}

Because of the limited software availability on the Snellius nodes, certain software versions are not identical when compared with our local node. The configuration of the local node can be seen in \autoref{tab:local_machine_config} and the Snellius node configuration in \autoref{tab:snellius_machine_config}. The performance impact of these version differences is minimal for our purposes, since our measurements concern GPU copy time and the architecture of external KV caches.

The SSDs on the test systems are not identical, and have slightly differing write performance. However, since our workload primarily stresses the read performance, this is not a problem. Further, the sustained read performance we measured for the offload drive used on each system, the Kioxia CM7-R locally and the Samsung PM1743 on Snellius, is similar at 13.5~GB/s.

\section{Characterization of Existing KV Cache Systems}
\label{sec:characterization}

To motivate and inform the design of our own external KV cache, \pykvcache{} (\autoref{sec:design}), we characterize existing KV cache implementations to identify sources of latency and which parts of their design can be improved. A particular focus was placed on the hot path to reduce TTFT, as well as overlap between copy and compute. In these experiments, the terms ``document length'' and ``prompt length'' are used interchangeably to refer to the number of tokens in a request.

Unless otherwise specified, all experiments in this section were performed on the local node using Llama~3.2~3B.

Unless otherwise specified, \offload{} refers to the default \vllm{} implementation of the \kvoffload{} for a CPU cache. All experiments in this section were run with GPU prefix caching disabled, to isolate CPU and disk effects. These experiments were run while \vllm{} was actively changing both cache interfaces, so they span several \vllm{} versions.

\subsection{GPU~$\leftrightarrow$~CPU Transfer Path}

Our first experiment aims to understand whether the choice between the \kvtransfer{} and \kvoffload{} has an impact on TTFT, and whether this effect changes with document length. We compare \lmcache{}, which uses the \kvtransfer{}, against \vllm{}'s default CPU cache implementation with the \kvoffload{}. We use the long-document benchmark over a range of document lengths, on \vllm{} \vllmVsixteen{}. Each run first sends $N$ unique prompts of length $D$ to populate the cache, then repeats each prompt $R$ times in random order, resulting in $N + (N \times R)$ requests. The first phase measures requests that compute and store KV data, while the second measures requests that load an existing prefix. All requests generate one output token so that the measurement is dominated by prefill or cache loading rather than decoding.

\begin{figure}[h]
  \centering
  \includegraphics[width=\columnwidth]{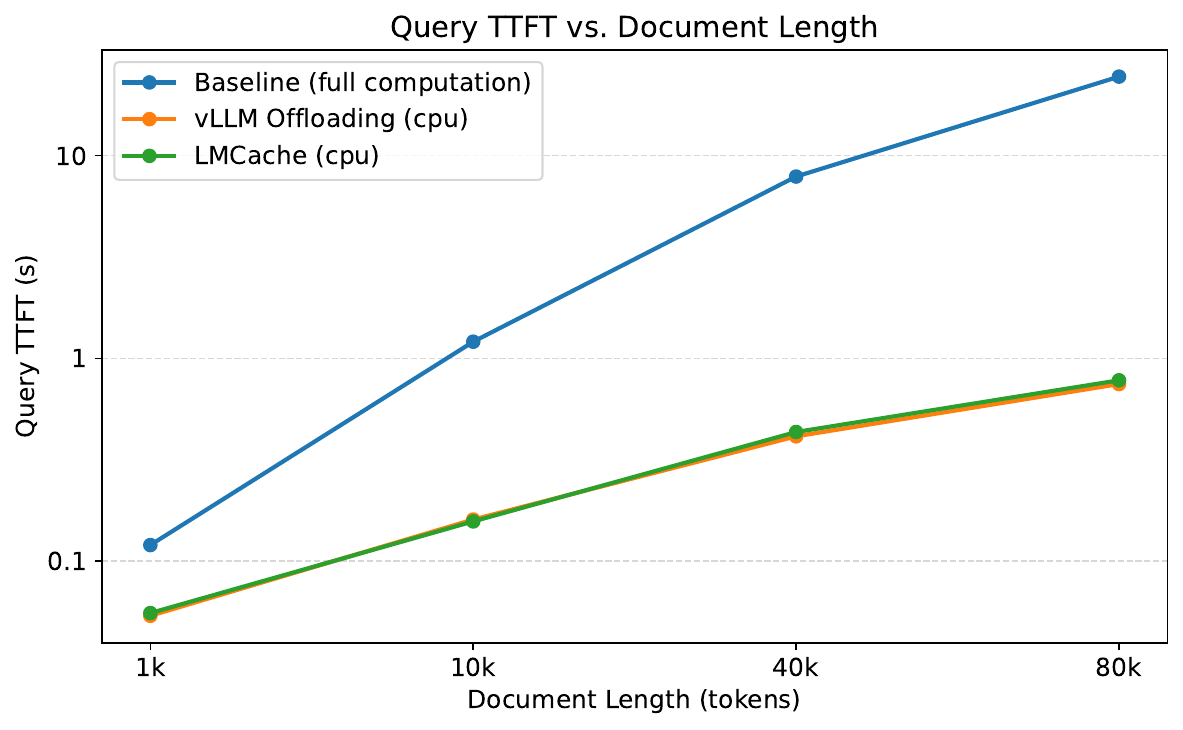}
  \caption{Long-document benchmark TTFT (\vllm{} v0.16, local node). Both external caches cut cache-hit TTFT by 2.2--32.8$\times$ relative to recomputation.}
  \label{fig:longdoc_ttft}
\end{figure}

\autoref{fig:longdoc_ttft} shows that both external caches substantially reduce cache-hit TTFT compared with recomputing the full prompt. The benefit grows with document length, from 2.2$\times$ at 1k tokens to 32.8$\times$ at 80k tokens, since the recomputation being replaced grows faster than the KV data being transferred. The two interfaces are far closer to each other than either is to the baseline. \offload{} is 1.03$\times$ faster at 1k tokens, 1.05$\times$ at 40k tokens and 1.04$\times$ at 80k tokens, while \lmcache{} is 1.02$\times$ faster at 10k tokens. In absolute terms these gaps are at most 33~ms. On this benchmark the choice of interface therefore has little effect on cache-hit TTFT, because the same KV data must cross the same CPU~$\rightarrow$~GPU path in both cases and that transfer dominates the measurement. This does not establish that the interfaces are equivalent in general, only that repeated long documents do not separate them. The next experiment therefore moves to shorter, mostly fresh traffic, where connector overhead is a larger fraction of TTFT.

The first experiment uses repeated long prompts, where a large transfer dominates every cache hit. We therefore use \textbf{\sharegpt{}} to test whether the two interfaces also match under shorter, mostly fresh traffic. We replay more than 10{,}000 \sharegpt{} prompts at 32 requests/s using the same two cache implementations. These results are not plotted. \lmcache{} has a TTFT of 206~ms, compared with 128~ms for the \offload{} implementation, a 1.61$\times$ difference. Because \sharegpt{} contains very little prefix reuse, this experiment does not measure the benefit of cache hits. Instead, it showcases connector overhead, and specifically focuses on the store path. This can affect TTFT even when most requests do not load a reusable prefix, and shows that the two interfaces do differ, on the store path rather than on the cache-hit path measured in \autoref{fig:longdoc_ttft}.

\subsubsection*{Tracing the GPU~$\leftrightarrow$~CPU Transfer Path}
To identify where the two connectors differ, we instrument both cache paths and record cache loads, stores, GPU~$\rightarrow$~CPU and GPU~$\leftarrow$~CPU copies, and model forward passes for a 40k prompt. The default \offload{} implementation uses asynchronous DMA copies between GPU and CPU memory. Enqueuing a copy through \texttt{transfer\_async} has negligible cost in our traces, whereas the corresponding \texttt{wait\_for\_save} call in \lmcache{} takes approximately 400~$\mu$s. \lmcache{} also performs its copies using a CUDA kernel, which consumes GPU compute resources.

\offload{} spent 3.67~s in total on GPU~$\leftrightarrow$~CPU copies across the benchmark and issued 21 transfers per warmup request, with each transfer taking an average of 9.08~ms. \lmcache{} spent 4.13~s on the same stage, but divided each request into 640 transfers averaging 310~$\mu$s each. \offload{} transfer size follows \vllm{}'s \texttt{max\_num\_batched\_tokens}, while \lmcache{} divides the same KV data into smaller cache chunks. The larger \offload{} transfers are interleaved with subsequent forward passes, allowing much of the copy time to overlap with model execution. This experiment therefore identifies transfer granularity and execution overlap as possible causes of the TTFT difference, but the different copy mechanisms mean that the aggregate timings alone cannot determine whether either system has higher raw transfer bandwidth.

The transfer trace shows that \offload{} copies overlap with later forward passes, but not why this overlap occurs. To determine the cause, we trace successive \vllm{} engine iterations while varying \texttt{max\_num\_batched\_tokens}. The scheduler greedily fills each execution batch up to this limit and submits the next iteration immediately after the current one. The \offload{} connector queues KV data produced in iteration $N$ and performs the corresponding store work during iteration $N+1$, allowing the copy to overlap with a later forward pass. The \transfer{} path used by \lmcache{} does not defer stores in the same way and instead waits for the store operations after each forward pass. This experiment confirms that the observed TTFT difference is partly a consequence of where store work is placed in the engine cycle, rather than only the cost of the copy itself.

To test whether the TTFT difference is caused by raw GPU~$\leftrightarrow$~CPU copy bandwidth, we vary both document length and \texttt{max\_num\_batched\_tokens} and calculate the effective bandwidth of the measured copies. This changes the size and number of transfers issued by the \kvoffload{} while keeping the underlying GPU and CPU unchanged.

\begin{figure}[h]
  \centering
  \includegraphics[width=\columnwidth]{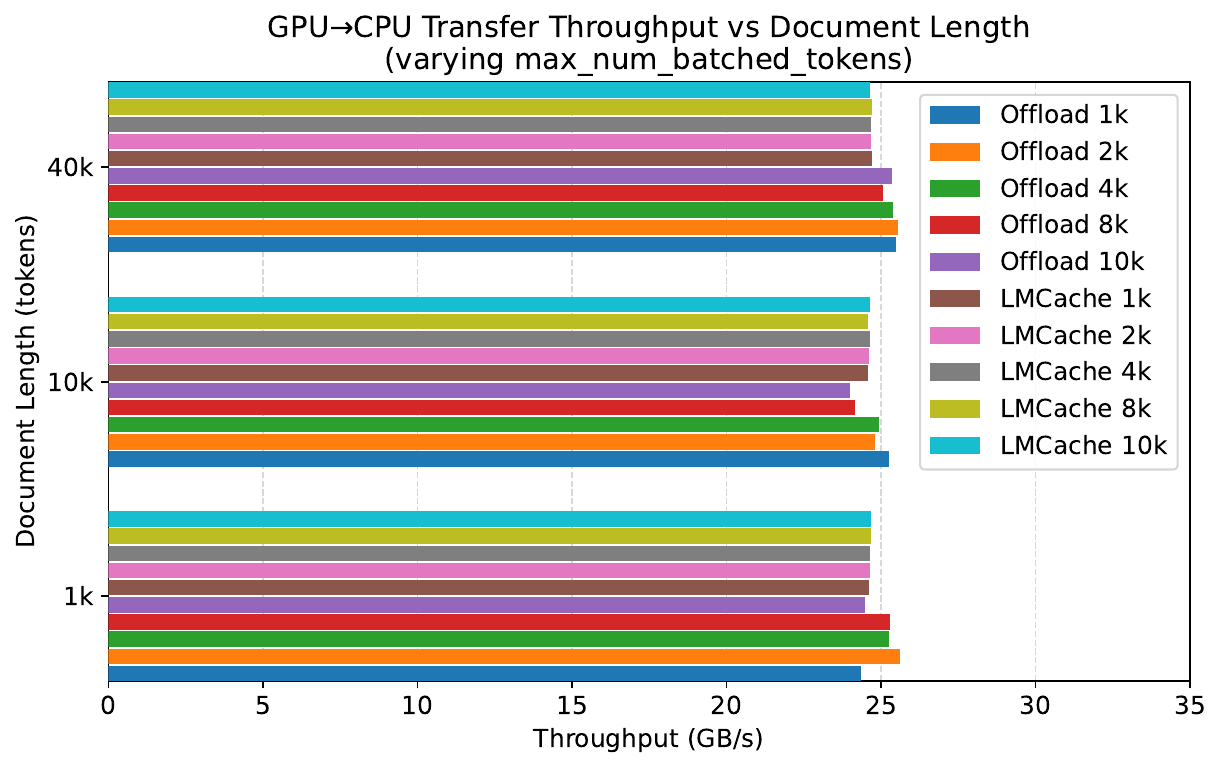}
  \caption{GPU~$\leftrightarrow$~CPU transfer bandwidth for different \texttt{max\_num\_batched\_tokens}. 1k--10k in the legend denotes the \texttt{max\_num\_batched\_tokens} value (\vllm{} v0.16, local node).}
  \label{fig:from_gpu_bw}
\end{figure}

Despite their different copy mechanisms and granularities, \autoref{fig:from_gpu_bw} shows that both systems sustain approximately 24--26~GB/s of GPU~$\leftrightarrow$~CPU bandwidth. Changing \texttt{max\_num\_batched\_tokens} has little effect on bulk transfer bandwidth once transfers are sufficiently large. We therefore conclude that the TTFT advantage of \offload{} comes mainly from issuing fewer transfers and placing them asynchronously alongside model execution, rather than from moving data at a substantially higher raw bandwidth. This motivates retaining \offload{}'s asynchronous execution model in \pykvcache{} while reducing coordination and I/O overhead elsewhere in the cache path.

\subsection{NVMe SSD KV Caching}
\label{subsec:nvme-kv-caching}

Having investigated the GPU~$\leftrightarrow$~CPU transfer path, we next examine the additional costs introduced when KV data is stored on an NVMe SSD. These experiments focus on the conditions under which disk KV caching is useful and on whether the overhead comes from the storage device, the filesystem, or the cache connector.

\begin{figure*}[t]
  \centering
  \includegraphics[width=\textwidth]{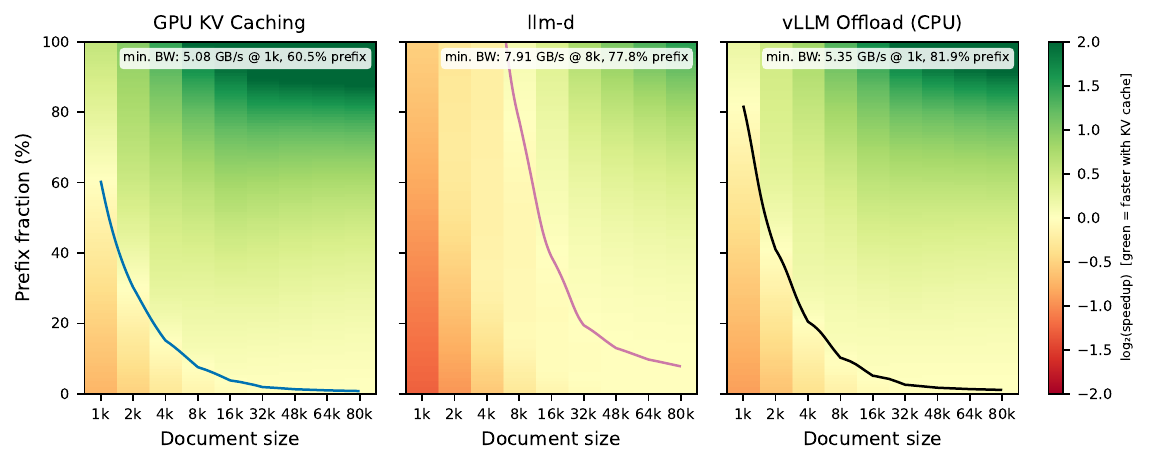}
  \caption{TTFT break-even frontiers (\vllm{} v0.22, Snellius node). Above each line, loading the cached prefix is faster than recomputing it.}
  \label{fig:other-kv-pareto}
\end{figure*}

Our first disk experiment identifies when loading a cached prefix is faster than recomputing it on the GPU. This is critical, since widely varying GPU and disk performance could swing results either way. Using the Pareto benchmark described in \autoref{sec:pareto}, we measure cold TTFT (compute + store op) and cache-hit TTFT (load op) across a sweep of document sizes. We then interpolate between the measured points to obtain the boundary at which both choices have equal TTFT. We validate this boundary by running additional configurations immediately above and below it. The experiments were conducted with \llmd{} as the filesystem KV cache, and as described in \autoref{subsec:llm-d}, the tests were run on our fork with direct I/O support to avoid page-cache effects. The frontiers discussed below were measured on the same \vllmVtwentytwo{} fork used in \autoref{sec:evaluation}.

The resulting Pareto frontiers in \autoref{fig:other-kv-pareto} show that a cache hit is not sufficient for caching to be beneficial. The line in each subplot marks the minimum reusable prefix fraction at which the TTFT for a cache hit will match recomputation, and the background shades the TTFT speedup or slowdown predicted for the same path against recomputing the whole prompt, at each document size and prefix fraction. On our Snellius node, using Llama~3.2~3B with \llmd{}, an 8k prompt requires a 77.8\% prefix reuse for breaking even, while at 80k tokens the break-even prefix falls to 7.8\% of the prompt. The frontier shifts between GPUs, models, SSD. Faster GPUs recompute a prefix more quickly, whereas larger or slower models leave more time in which cached KV data can be loaded.

\begin{figure}[t]
  \centering
  \includegraphics[width=\columnwidth]{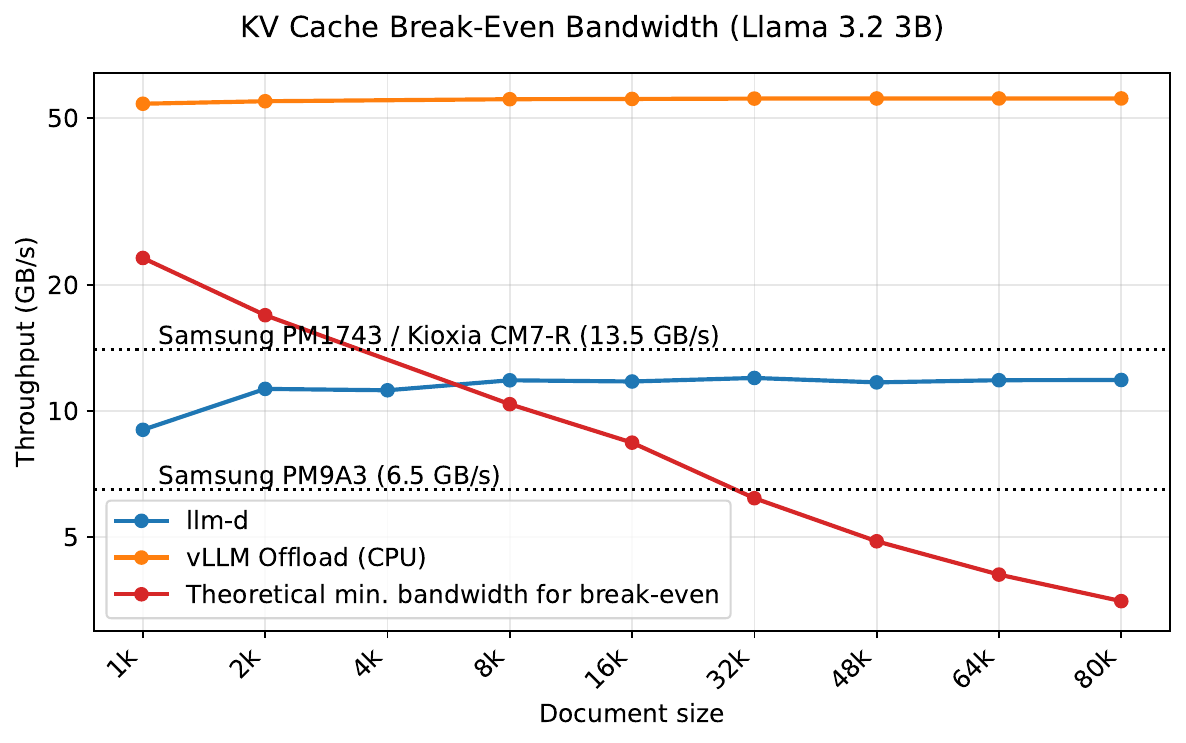}
  \caption{Break-even storage bandwidth (\vllm{} v0.22, Snellius node). This is a bandwidth representation of the Pareto Frontiers in \autoref{fig:other-kv-pareto}. The dashed lines show ideal measured throughput of our SSDs.}
  \label{fig:min-bandwidth}
\end{figure}

In \autoref{fig:min-bandwidth} we answer the question of how fast a storage platform (DRAM, SSD, etc.) must be for caching to beat recomputation. Based on \autoref{sec:pareto}, for a document of $D$ tokens we subtract the non-transfer part of the cache hit from the time needed to compute that document, which leaves the time budget available to move the KV data:
\[
  \mbox{max\_io}(D) = f(D) - \left(g(D) - t_{\mbox{\scriptsize copy}}(D)\right),
\]
where $g(D)$ is the cache-hit TTFT and $t_{\mbox{\scriptsize copy}}(D)$ is the time spent moving the KV data into the GPU, taken from the profiler traces described in \autoref{subsec:tracing} as the median duration of the recorded transfer events for that path. In \llmd{} this is disk to GPU time, and covers the disk read as well as the GPU copy. Dividing the document's KV bytes by this budget gives the throughput a cache path must sustain for a fully cached prompt to match recomputing it. Llama~3.2~3B stores 114{,}688~bytes per token, so the requirement falls sharply as prompts grow, from 23.2~GB/s at 1k tokens to 10.4~GB/s at 8k and 3.5~GB/s at 80k, because prefill cost grows faster than the KV data that prefill produces.

The GPU~$\leftrightarrow$~CPU copy sustains 54--56~GB/s and therefore clears the requirement at every document size. \llmd{} moves 9.0~GB/s at 1k tokens and 11.2 to 12.0~GB/s at larger sizes, which leaves it below the requirement at 1k and 2k but above it from 8k onward. The reference lines represent the ideal throughput of drives in our setup, as well as the required throughput for break-even. The practical consequence is that storage bandwidth sets a minimum document size. Below roughly 8k tokens the prefill is short enough that no storage device we measured can load the prompt faster than the GPU rebuilds it, and on a mid-range drive that threshold moves out to 32k. Therefore, external cache admission must depend on the model, GPU, SSD, and reusable prefix length rather than treating every hit as beneficial.

\begin{figure*}[t]
  \centering
  \begin{subfigure}[t]{0.48\textwidth}
    \centering
    \includegraphics[width=\linewidth]{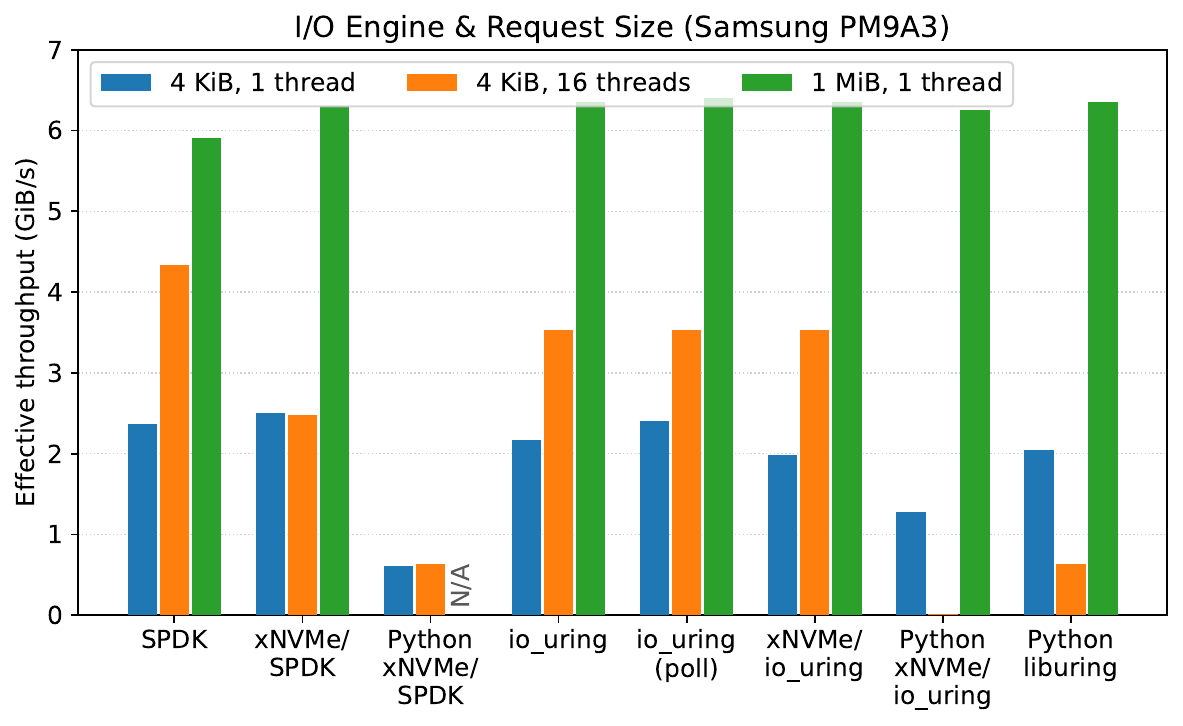}
    \caption{I/O-engine throughput.}
    \label{fig:io-engine-request-size}
  \end{subfigure}
  \hfill
  \begin{subfigure}[t]{0.48\textwidth}
    \centering
    \includegraphics[width=\linewidth]{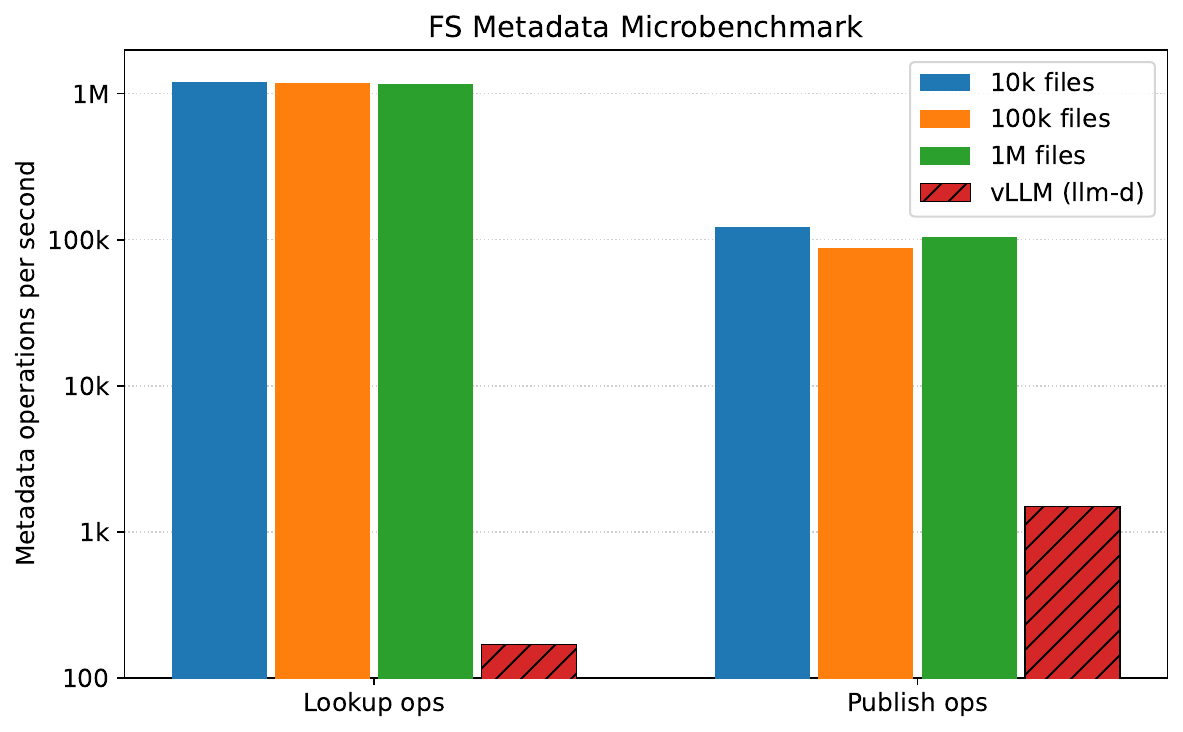}
    \caption{Filesystem metadata throughput.}
    \label{fig:fs-metadata}
  \end{subfigure}
  \caption{Storage microbenchmarks.}
  \label{fig:storage-microbenchmarks}
\end{figure*}

The Pareto model predicts the best TTFT attainable from the measured compute and transfer costs. Initial \llmd{} load measurements remained slower and more variable than this prediction even when disk reads approached the raw throughput measured by \fio{} (v3.36) \cite{fio}. To identify the source of this gap, we instrumented the \llmd{} filesystem connector and traced its memory copies and block I/O requests.

The trace showed that \llmd{} copied and submitted data separately for each cache chunk. More importantly, its minimum staging-buffer allocation was larger than the KV data required by our test model. A request containing 4.4~GB of useful KV data consequently caused approximately 10~GB to be copied and read or written. We reported this allocation problem, and a separate \vllm{} integration incompatibility, upstream as \llmd{} issues \href{https://github.com/llm-d/llm-d-kv-cache/issues/454}{\#454} and \href{https://github.com/llm-d/llm-d-kv-cache/issues/389}{\#389}; the allocation bug was fixed in \href{https://github.com/llm-d/llm-d-kv-cache/pull/589}{\#589}. After removing the minimum allocation, TTFT fell approximately 50\% below \lmcache{} in the same test. The connector's reads could reach more than 12~GB/s, but writes remained around 2~GB/s. Device bandwidth alone is therefore an incomplete predictor of cache performance: staging-buffer policy and data amplification can dominate the storage path. This motivates exact staging allocations and explicit bounds on how much intermediate memory an operation may reserve.

A separate native \vllm{} offloading issue observed during \scbench{} is discussed alongside its execution trace in \autoref{fig:pool-compare}.

\subsubsection*{I/O engine and request size}
We next test whether a more scalable asynchronous I/O engine or a larger worker pool improves SSD KV caching on our local node. As reference points, \fio{} measured 13.5~GB/s on the Kioxia drive, and 6~GB/s from the Samsung PM9A3, while \texttt{nvbandwidth} measured 25--26~GB/s between CPU and GPU memory on our local node, connected via a 16$\times$ PCIe 4.0 link. Our Snellius node measured 54--56~GB/s over a 16$\times$ PCIe 5.0 link. In a standalone storage benchmark on Snellius, \iou{} reached 338k operations/s with one thread, compared with 13k operations/s for the tested POSIX path, and reached 2.59 million operations/s with 16 threads. However, replacing the I/O path did not measurably improve TTFT in the LLM benchmark. Block tracing explains this result: the filesystem issued mostly 512~KiB requests, with some 1~MiB requests. As shown in \autoref{fig:io-engine-request-size} the tested engines vary considerably on 4~KiB requests, but converge near the drive's bandwidth limit with 1~MiB requests and a single thread.

The figure compares \spdk{}'s userspace NVMe driver, native \iou{} with and without submission-queue polling, the \xnvme{} wrappers over both backends, and the Python bindings for \xnvme{} and for \liburing{}. With one thread at 4~KiB the engines land close together, but at 16 threads only the native engines gain throughput while the Python paths lose it, which we attribute to the Python runtime rather than to a property of the device. At 1~MiB the difference disappears and every engine clusters near the 6~GB/s that \fio{} measured on this drive, except the Python \xnvme{} \spdk{} binding, which did not run at that request size.

The LLM workload is therefore limited by bandwidth and the placement of large transfers rather than small-I/O operation rate. High small-I/O IOPS and additional worker threads can improve a synthetic microbenchmark without improving end-to-end inference. This finding motivates using a small number of asynchronous workers with bounded queue depth, rather than relying on a large thread pool to obtain storage parallelism. Since KV cache files are much larger than 1~MiB, a Python implementation will reach the same storage throughput as a native one (\autoref{sec:design}).

\subsubsection*{Filesystem metadata scalability}
Because \pykvcache{} is intended to use an ordinary filesystem and to be shared by multiple \vllm{} instances, we test whether filesystem metadata becomes a bottleneck as the cache grows. We populate workspaces containing 10k, 100k, and one million files and separately measure the lookup and publication transactions defined in \autoref{sec:fs_metadata}. As summarized in \autoref{fig:fs-metadata}, lookup throughput remains nearly constant at 1.21, 1.19, and 1.17 million operations/s, respectively. Publication operations achieve 122.7k, 87.4k, and 104.5k operations/s.

For comparison, the instrumented \vllm{} workloads issue only approximately 170 lookups/s and 1{,}500 publications/s. A standalone connector test with up to 64 instances likewise showed no filesystem-specific performance collapse before CPU capacity became the limiting factor. We conclude that a hierarchical filesystem layout can support the metadata rate required by the evaluated serving workloads. This does not make individual metadata operations free, but it shows that the filesystem namespace itself is unlikely to be the throughput bottleneck at the tested scale.

\subsubsection*{Placement of disk reads}
We finally use the request traces to determine whether disk I/O overlaps with useful work or remains on the cache-hit critical path. In the existing disk paths, a request waits for the disk $\rightarrow$ CPU transfer before the remaining CPU $\rightarrow$ GPU transfer can be performed. Thus, even when the SSD reaches its expected bandwidth, the read remains visible in TTFT if it begins only after the request enters execution. This observation motivates investigating whether disk reads for queued requests can begin earlier i.e., through preloading.

\subsubsection*{Characterization Takeaway}

These experiments identify the requirements carried into the design of \pykvcache{}. The cache should avoid loading or storing prefixes below the measured break-even point, and allocate staging memory in proportion to the useful KV data. Its disk path should target large, bandwidth limited I/O with bounded concurrency rather than trying to achieve high IOPS with numerous threads and small requests. A regular hierarchical filesystem is sufficient for the observed metadata load, and disk reads should begin before request execution if possible so that storage latency does not remain entirely on the TTFT critical path. The next section describes how \pykvcache{} implements these requirements.

\section{Design of py-kvcache}
\label{sec:design}

The characterization of existing KV cache systems identified various problems. Based on these findings, we built \href{https://github.com/atlarge-research/py-kvcache/tree/790074addbccbe95eda5bc4cfe6afe71d524a910}{\pykvcache{}}, a Python \vllm{} \kvoffloadname{} connector for shared filesystem storage, built on the \kvoffload{} and its asynchronous GPU transfers. The design has four goals: to share cached prefixes across \vllm{} instances through an ordinary filesystem layout; to reach the bandwidth of current-generation NVMe devices without a large I/O thread pool; to bound the intermediate CPU memory any operation may reserve; and to avoid cache operations that are unlikely to improve latency.

\subsection{Selecting the Storage Interface}

The final I/O path was reached through several prototypes. We first investigated \xnvme{} because it provides a common interface to Linux I/O engines and the \spdk{} userspace NVMe driver. \spdk{} was attractive because bypassing the kernel storage stack could reduce per-operation overhead. However, the \xnvme{} \spdk{} backend uses only \spdk{}'s userspace NVMe driver, rather than its reactor, threading model, or application framework \cite{xnvmeSpdk}. It also requires the device to be detached from the kernel NVMe driver and accessed by its PCI address and namespace. This did not fit our goal of allowing multiple cache instances to exchange hash-addressed objects through a mounted filesystem.

The prototype also exposed practical API problems. Our first \xnvme{} implementation aimed to use \iou{}, but the \xnvme{} file API ignored this setting and executed the operations through POSIX I/O. After moving to the asynchronous \xnvme{} interface, each file operation required its own buffer and ring. The \spdk{} path additionally required the NVMe command interface, making it different from the filesystem backends, while the \xnvme{} \spdk{} integration did not expose all of the functionality needed by the prototype. Correcting the backend improved storage microbenchmarks (\autoref{fig:io-engine-request-size}), but did not improve our \vllm{} TTFT. This result agreed with our prior characterization, since the workload consists of large, bandwidth limited reads and writes, so reducing overhead of I/O operations does not necessarily shorten the request's critical path. Further, this initial implementation was only able to maximize disk throughput in a traditional multithreaded worker design, similar to \llmd{}. In a single threaded design, the overhead of maintaining individual rings and separate polling queues for each file led to significant overhead for a single Python thread and as a result we could not utilize all of the available disk throughput.

We separately tested whether GPUDirect Storage could remove the intermediate CPU transfer. We implemented a simple prototype using KvikIO, which provides Python bindings to \texttt{cuFile} and can access GPU buffers through GPUDirect Storage \cite{kvikio}. In our configuration, this path was slower than all our other implementations.

\pykvcache{} therefore uses ordinary files with direct I/O and \iou{} through the \liburing{} Python library (v2026.3.30) \cite{liburing}. \iou{} provides asynchronous submission and completion queues shared between userspace and the kernel, and permits several requests to be submitted together \cite{ioUring}. This preserves our need for asynchronous I/O and bounded queue depth while also allowing the implementation to remain in Python, because at the request sizes used for KV blocks, we were easily able to maximize the throughput available on our test setups.

\begin{figure*}[t]
  \centering
  \includegraphics[width=0.7\textwidth]{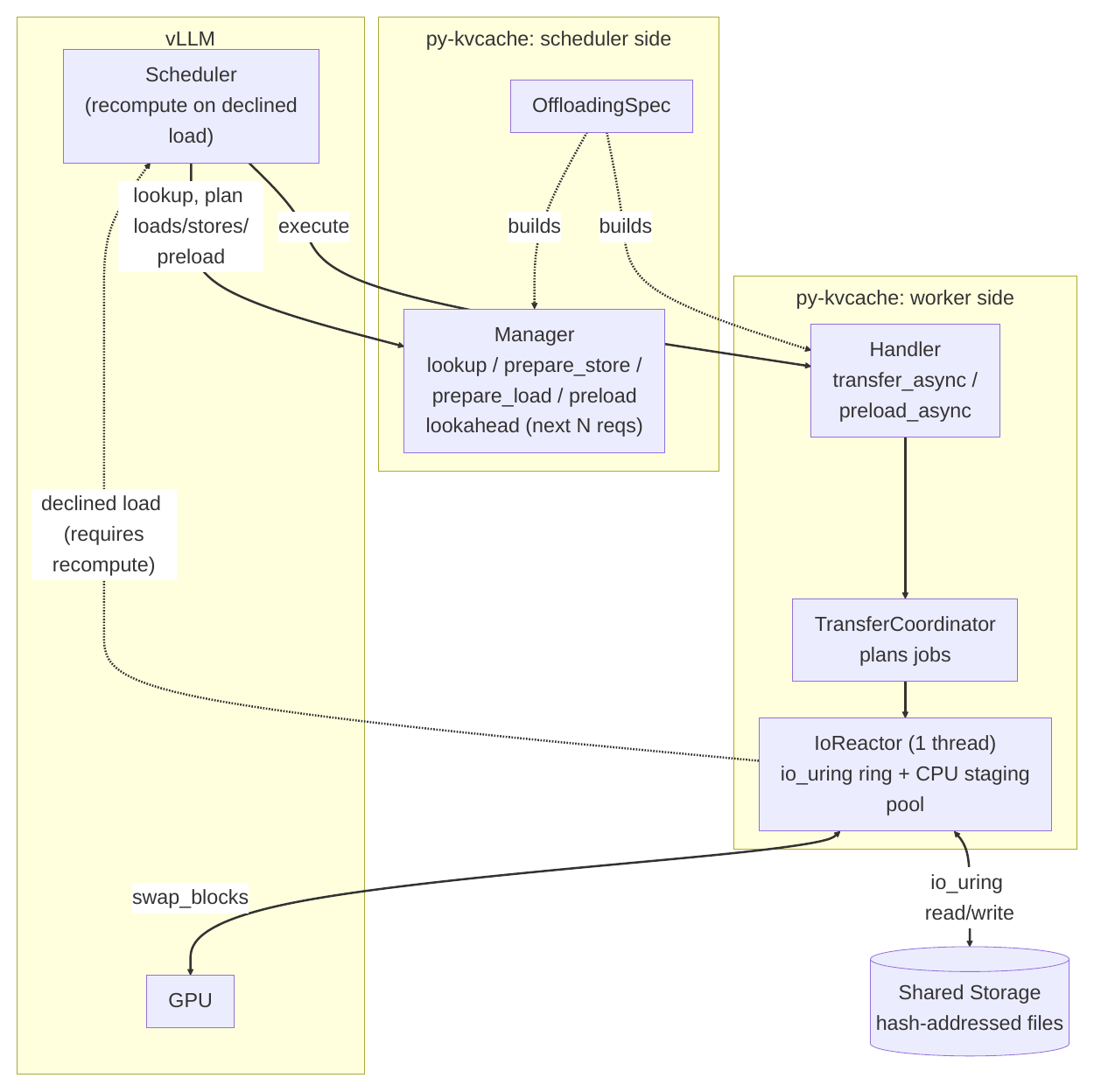}
  \caption{\pykvcache{} architecture.}
  \label{fig:arch}
\end{figure*}

\subsection{System Architecture}

\autoref{fig:arch} shows the resulting architecture. Solid arrows show request and KV-transfer execution, while dotted arrows show construction and fallback control paths. \pykvcache{} is loaded by \vllm{} as an \texttt{OffloadingSpec}, which constructs components in both the scheduler and worker processes. On the scheduler side, the \texttt{Manager} performs cache lookups and prepares load, store, and preload plans. It can decline a load, in which case \vllm{} recomputes the corresponding prefix, but it does not move KV data itself. On the worker side, the \texttt{Handler} receives the plan and invokes the \texttt{TransferCoordinator}. The coordinator divides the operation into file jobs, and a single \texttt{IoReactor} schedules them subject to the available staging memory and I/O depth. The reactor owns the \iou{} queues, CPU staging pool, and CUDA streams, and uses \vllm{}'s batched block-copy operation to move data between the GPU KV tensors and staging slots.

This separation keeps scheduling decisions close to \vllm{}'s request state and storage execution close to the model worker. It also makes the control path independent of the data path: only compact transfer plans cross between the scheduler and worker, while KV tensors move directly between the worker's GPU, its staging pool, and shared storage.

\subsection{Shared Filesystem Layout}

To share cached prefixes across \vllm{} instances, \pykvcache{} adopts the hierarchical hash-addressed layout used by \llmd{}. A cache entry's path is derived from the model configuration and prefix-block hash, and intermediate directories distribute entries across the namespace. Our metadata characterization showed that this organization supports substantially more lookups and publications than the serving workload requires, including with one million populated files.

File data is accessed using direct I/O so that transfers do not depend on the state of the OS page cache. Direct I/O alone does not make an entry safe to share. Stores therefore write to a uniquely named temporary file in the target directory. After the write completes, \pykvcache{} creates the final hash-addressed name with a hard link and removes the temporary name. Link creation is atomic and fails if another instance has already published the same hash \cite{posixLink}, giving concurrent writers a first-writer-wins protocol without replacing a valid entry. Readers only open the final name, so an entry is either absent or complete.

\subsection{Asynchronous Transfers and Bounded Staging}

\pykvcache{} stores KV data in blocks that group several of \vllm{}'s GPU blocks, with one file per block. With the 256-token block size used in our experiments, one Llama~3.2~3B cache block occupies approximately 28~MiB. Existing filesystem backends use pools of blocking worker threads to obtain I/O concurrency, which is not beneficial at this object size. \pykvcache{} instead uses one reactor thread to submit and reap multiple outstanding operations. The configured I/O depth, rather than the number of worker threads, controls storage concurrency.

By having a configurable software I/O depth, we can bound memory allocation. A single load or store covers the whole prefix of one request, which usually spans many storage blocks. \pykvcache{} therefore splits it into one job per storage block, each of which occupies one slot of a fixed CPU staging pool, and returns the slot once its I/O and GPU transfer complete. This minimizes CPU memory pressure and avoids deadlocks or similar issues that could arise when different scheduling is used, e.g., through deferral of requests. A request larger than the pool is processed incrementally rather than reserving memory for the entire KV object. Loads, stores, and preloads use the same pool, preventing each path from independently reserving its required capacity. The staging pool is one contiguous aligned, pinned CPU allocation. The CPU pool can optionally retain completed slots with LRU or ARC replacement, making it a DRAM cache in addition to a transfer buffer. By using the same CPU slots for disk I/O and DMA copies to our GPU, we avoid issues such as the write amplification observed in \llmd{} (\autoref{subsec:nvme-kv-caching}).

Transfers are pipelined per storage block (i.e. each file) rather than executing as a single operation. On a load, the reactor first issues asynchronous \texttt{openat} operations through \iou{} up to a separate lookahead depth. A slot is acquired only when I/O depth is available and the read can be submitted. When a disk read is completed, its block mapping is immediately queued for CPU$\rightarrow$GPU transfer. All load mappings that become ready during one reactor iteration, including mappings from different requests, are fused into a single \texttt{swap\_blocks\_batch} launch on a dedicated CUDA stream. While that DMA is in flight, the reactor can reap other disk completions and submit more reads. The staging slot is released only after its CUDA event reports completion. This pipelining can be observed in the trace in \autoref{fig:trace-sample}. The pool reserves additional copy headroom beyond the configured I/O depth so that slots held by CUDA transfers cannot drain the disk pipeline.

Stores use the reverse pipeline. Each staging slot has its own CUDA stream, which waits on that slot's CUDA event and copies one storage block from GPU memory without a device-wide synchronization. As soon as a slot's GPU$\rightarrow$CPU copy completes, the reactor submits its I/O write. A store consumes one I/O-depth budget entry from copy launch until the write completion is reaped. The completed file is then published, and the slot is released or retained in the optional DRAM cache. Together with \offload{}'s deferred store semantics, this permits GPU copies and disk writes from one engine iteration to overlap work in later iterations, although this was never observed in our experiments because of our high performance NVMe drives, and was also implicitly limited by \texttt{max\_num\_batched\_tokens}.

\subsection{Preloading}
\label{subsec:preloading}

Although asynchronous I/O reduces blocking inside the transfer engine, an ordinary demand read still begins after the request is selected for execution. \pykvcache{} therefore uses scheduler information to preload queued requests. The manager examines a bounded lookahead of waiting requests and sends preload plans to the worker when the storage path has available capacity. The reactor begins moving matching prefixes from disk into the existing CPU staging pool before those requests are scheduled. If a request later begins execution, it joins an in-flight read or claims the staged blocks, leaving only CPU $\rightarrow$ GPU promotion on its request path. This can lead to significant reductions in TTFT.

This required a small set of changes to our \href{https://github.com/t348575/vllm/tree/d6eadf416bb5234047760bf55d532f2f038cf697}{\vllm{} fork}. The original \kvoffload{} only asks the worker to load KV data after the scheduler has selected a request, while the worker itself has no visibility into the waiting queue. We added a bounded \texttt{on\_preload\_candidates} callback to the scheduler and propagated the resulting preload identifiers and block metadata to the worker, where a new connector hook can start the read. This scheduler information stays within \vllm{}, and is handled in the translation layer between the \kvtransfer{} and the \kvoffload{}. The later demand load identifies and claims the same staged operation. Without this scheduler-worker path, \pykvcache{} could not begin disk I/O before request admission, and the full read would remain on the TTFT critical path.

\begin{figure*}[t]
  \centering
  \begin{subfigure}[t]{0.49\textwidth}
    \centering
    \includegraphics[width=\linewidth]{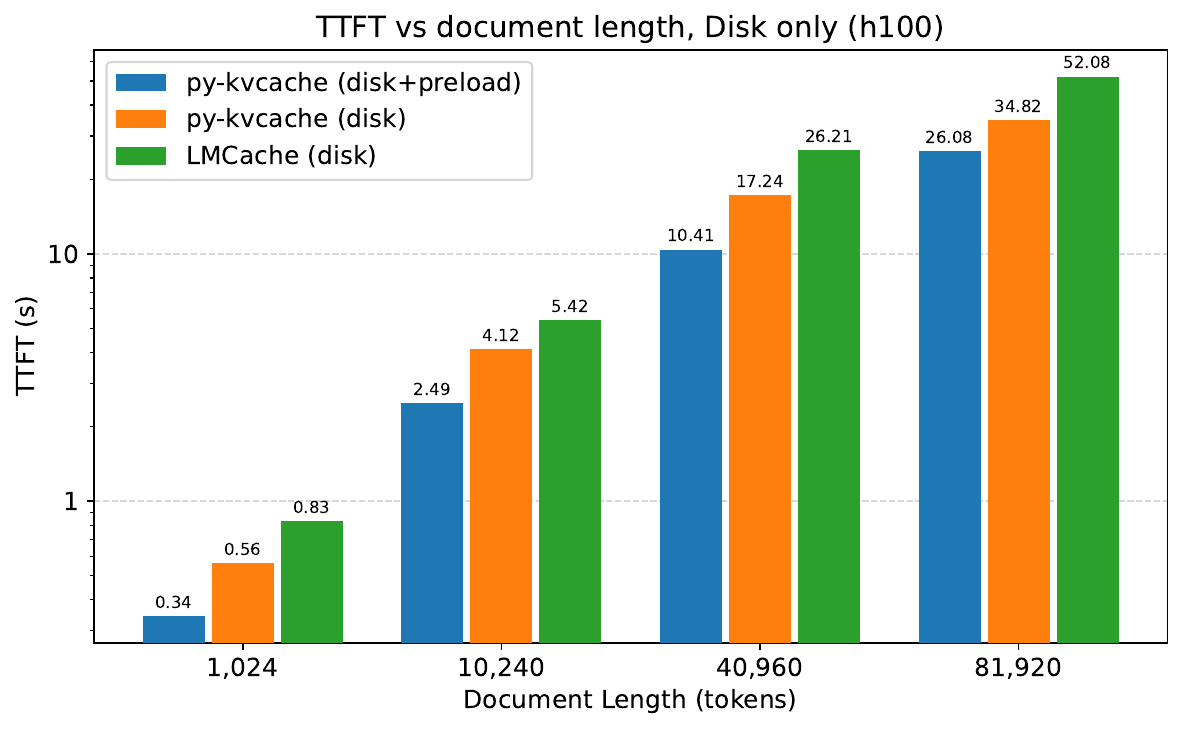}
    \caption{Disk-only query TTFT .}
    \label{fig:disk-only-ttft}
  \end{subfigure}
  \hfill
  \begin{subfigure}[t]{0.49\textwidth}
    \centering
    \includegraphics[width=\linewidth]{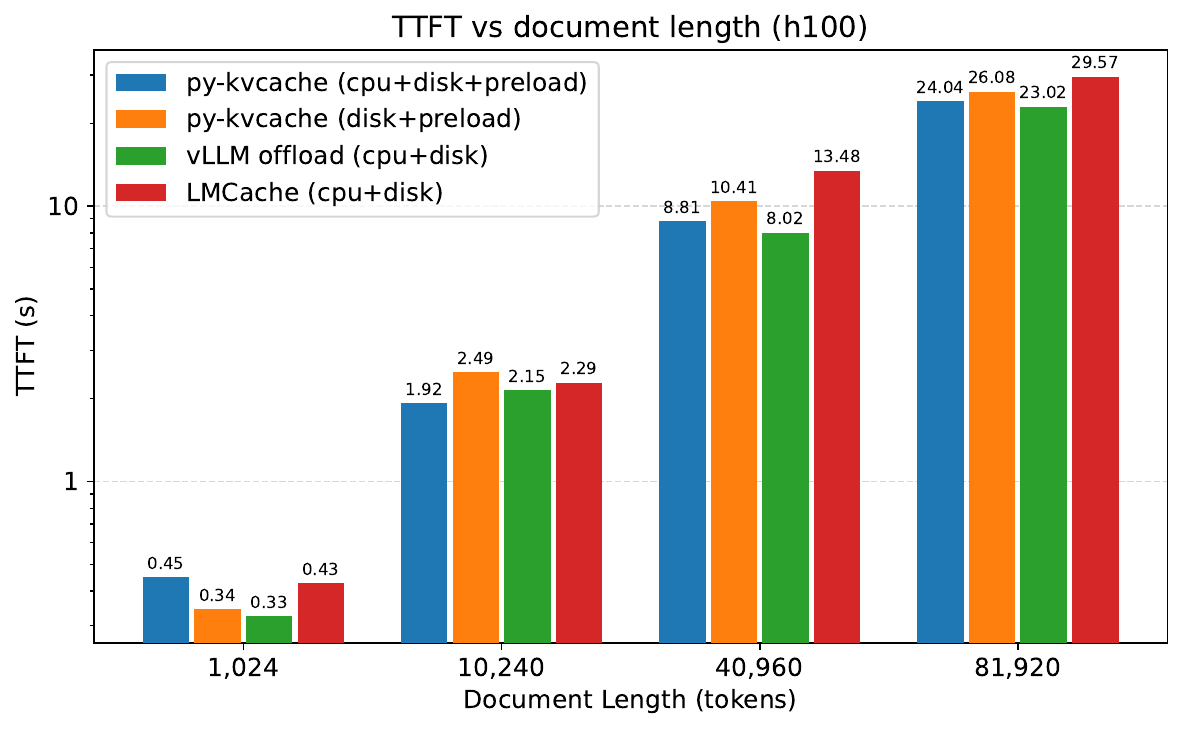}
    \caption{Disk + DRAM query TTFT.}
    \label{fig:tiered-ttft}
  \end{subfigure}
  \caption{Isolated long-document evaluation (\vllm{} v0.22, Llama~3.2~3B, Snellius node).}
  \label{fig:controlled-evaluation}
\end{figure*}

Preloading work is deliberately subordinate to demand traffic. The reactor schedules ready demand read ops first, then new demand load and store ops, and only then speculative preloads. It reserves a number of staging slots equal to the configured I/O depth for foreground work, and a foreground load may reclaim either a retained cache slot or an unclaimed preload slot. If a demand load op appears after a speculative file has been opened or is in progress, the reactor closes that descriptor and requeues the preload instead of letting it consume read bandwidth. Identical prefixes requested by multiple preload candidates share one disk read and one reference-counted staging slot, while later demand loads join the in-flight operation rather than issuing duplicate reads. These rules were introduced after observing that concurrent speculative promotions in \vllm{}'s native secondary tier (\autoref{subsec:native-offload}) could exhaust CPU memory, evict recently promoted blocks, and force the scheduled request to recompute them, which we trace in \autoref{fig:pool-compare}.

\subsection{Break-even}

Finally, \pykvcache{} does not treat every matched prefix as a useful load. The Pareto front procedure described in \autoref{sec:pareto} is run once per node and model as an offline calibration, and produces a small file holding the break-even prefix length for each source tier. Nothing is measured or fitted while serving. At lookup, the manager compares the reusable prefix with the threshold for its source tier. If loading is predicted to cost more than recomputation, the manager declines the load and lets \vllm{} recompute the prefix. The gate therefore protects against a loss rather than producing a gain. Below the break-even point the prompt is short enough that its TTFT is small in absolute terms, so declining the load avoids wasted transfer work without materially improving latency. Its value is that it keeps external caching from becoming a regression on workloads such as the \bailian{} traces.

\section{Evaluation of py-kvcache}
\label{sec:evaluation}

We evaluate whether \pykvcache{} improves end-to-end serving performance, which parts of its design provide observed improvements, and whether its behaviour remains useful under realistic long-context workloads. We first use the controlled long-document workload to isolate disk and preload behaviour, then use \longbench{}, \scbench{}, and the \bailian{} traces to evaluate the complete system. Experiments in this section were performed on our local node and on the Snellius node, with Llama~3.2~3B or Qwen3~4B. All measurements in this section use \vllm{} \vllmVtwentytwo{}. \llmd{} is not evaluated here, since it was deprecated during this work in favour of \vllm{}'s native filesystem tier, which is functionally identical (\autoref{subsec:llm-d}).

\subsection{Isolated prefix cache tests}

\subsubsection*{Disk-only comparison}
We first ask whether \pykvcache{} improves cache-hit latency when all reusable KV data must be obtained from disk. We compare \lmcache{}'s disk backend with \pykvcache{}, both with and without preload, using 50 concurrent requests on the Snellius node. GPU prefix caching and the break-even gate are disabled so that every matched prefix exercises the disk path.

Across the tested document sizes, the query TTFT results in \autoref{fig:disk-only-ttft} show that \pykvcache{} without preload is consistently about 1.5$\times$ faster than \lmcache{}, isolating the benefit of its asynchronous transfer engine. With preload enabled the total speedup over \lmcache{} reaches 2.0--2.5$\times$; \autoref{subsec:preload-effect} separates the two contributions.

\begin{figure*}[t]
  \centering
  \includegraphics[width=\textwidth]{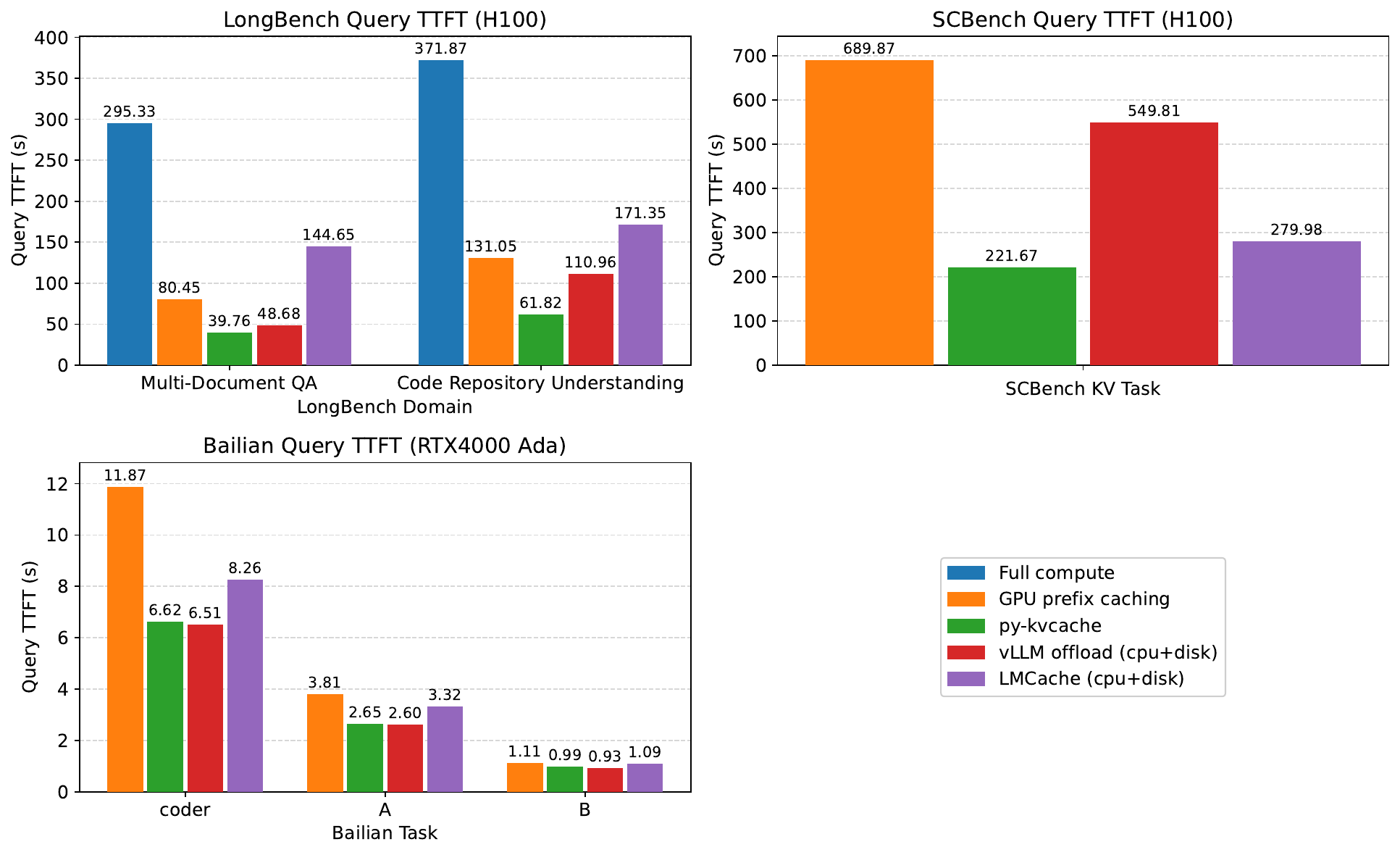}
  \caption{Query TTFT for \longbench{} and \scbench{} workloads on the Snellius node, and \bailian{} traces on the local node (\vllm{} v0.22, Qwen3~4B, Snellius node). \pykvcache{}, \lmcache{}, and native \vllm{} \kvoffloadname{} run with GPU prefix caching enabled.}
  \label{fig:longbench-ttft}
  \label{fig:scbench-ttft}
  \label{fig:bailian-ttft}
  \label{fig:trace-workloads}
\end{figure*}

\subsubsection*{Tiered KV caches}
We next compare complete ``production-like'' configurations with CPU DRAM, disk. As shown in \autoref{fig:tiered-ttft}, \pykvcache{} is 1.19$\times$, 1.53$\times$, and 1.23$\times$ faster than \lmcache{} at 10k, 40k, and 80k tokens respectively. It stays within 1.10$\times$ and 1.04$\times$ of the native \vllm{} offloading implementation at 40k and 80k tokens, and is 1.12$\times$ faster than it at 10k. Relative to the same \pykvcache{} configuration without a CPU tier, adding DRAM improves TTFT by 1.30$\times$ at 10k tokens, 1.18$\times$ at 40k, and 1.08$\times$ at 80k. At 1k tokens all four configurations fall within 0.12~s of each other and the ordering reverses, which is expected below the break-even point established in \autoref{subsec:nvme-kv-caching}. Thus, \pykvcache{} approaches the integrated native \vllm{} implementation at the document sizes where external caching is worthwhile.

\begin{figure*}[t]
  \centering
  \includegraphics[width=\textwidth]{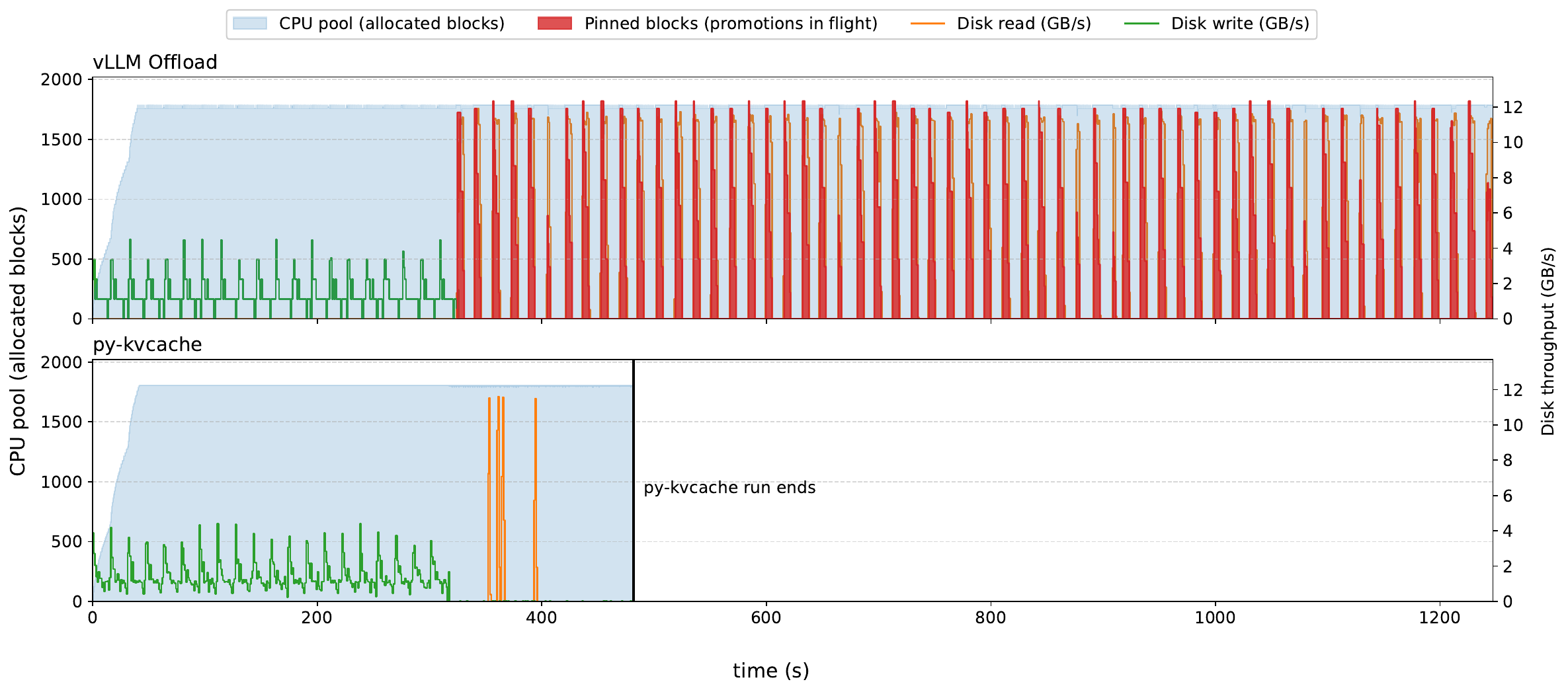}
  \caption{\scbench{} CPU-pool occupancy and storage throughput (\vllm{} v0.22, Qwen3~4B, Snellius node). Unbounded promotions keep the native CPU pool full and read 3.4~TB from disk, whereas \pykvcache{}'s bounded staging reads 85~GB and finishes serving all requests by 480~s.}
  \label{fig:pool-compare}
\end{figure*}

\subsection{Effect of Preloading}
\label{subsec:preload-effect}

The preceding results compare complete systems, so we next isolate our preload mechanism using identical disk-only \pykvcache{} configurations with preload enabled and disabled. The query TTFT results in \autoref{fig:disk-only-ttft} show that preload provides a 1.66$\times$ improvement at 40k tokens and a 1.34$\times$ improvement at 80k tokens. The smaller gain at 80k tokens is because the CPU $\rightarrow$ GPU transfer is significantly larger than at 40k tokens, and this time dominates the total TTFT, effectively ``hiding'' our preload gains.

Separately, we run a mixed workload containing 50 requests of 80k tokens, with 50\% of requests reusing a cached prefix and a maximum concurrency of eight. For this complete benchmark run, preload reduces the query-round wall time by 6.8\%, from 73 to 68~s. Within the same run, the per-request mean TTFT falls by 0.4~s across all requests and by 1.3~s for prefix-reuse requests alone. Varying the output length between 128 and 1k tokens produces little change in the preload benefit, as expected because the mechanism affects prefill rather than the subsequent decode phase.

Tracing explains where the improvement originates. Without preload, the first file read begins approximately 6~ms after cache lookup starts, after which the request must still wait for the disk $\rightarrow$ CPU transfer followed by a CPU $\rightarrow$ GPU transfer. Preload pays the same I/O startup cost earlier, while the request is waiting in the scheduler. When the request enters execution, some or all of the disk stage has already completed.

\subsection{Representative Long-context Workloads}

\subsubsection*{LongBench}
The controlled workload always constructs reusable prefixes, so we next test whether the same advantage appears in a less regular request sequence. We replay the ``Multi-Document QA'' and ``Code Repository Understanding'' domains from \longbench{} on the Snellius node. As shown in \autoref{fig:longbench-ttft}, \pykvcache{} achieves the lowest mean query TTFT in both domains: it is 6.02--7.43$\times$ faster than recomputation, 2.02--2.12$\times$ faster than GPU prefix caching, 2.77--3.64$\times$ faster than \lmcache{}, and 1.22--1.79$\times$ faster than the native \vllm{} \kvoffloadname{} implementation. Unlike the controlled benchmark, these workloads contain prefix chains of different lengths coupled with high concurrency. The result shows that preloading and \pykvcache{} as a whole is very effective in more realistic workloads where the reusable prefix distribution is not uniform. The performance uplift \pykvcache{} offers is largely due to where the transfer is placed. Requests queue behind one another at this concurrency, and \pykvcache{} preloads from the scheduler's waiting list rather than on a cache lookup (\autoref{subsec:preloading}), so a reusable prefix can be read while its request is still waiting, leaving only the CPU $\rightarrow$ GPU promotion on the critical path.

\subsubsection*{SCBench}
We use the \scbench{} KV workload to test a larger multi-turn working set in round-robin order. It is similar to our \longbench{} workloads, but represents a worst case scenario for KV caching. As shown in \autoref{fig:scbench-ttft}, \pykvcache{} achieves a 3.11$\times$ speedup over GPU prefix caching, a 2.48$\times$ speedup over native CPU-and-disk offloading, and a 1.26$\times$ speedup over \lmcache{} on the Snellius node.

The unexpectedly poor native-offload result led us to trace promotions, stores, demand loads, memory reservations, and evictions. \autoref{fig:pool-compare} compares native \vllm{} filesystem offloading in the top panel with \pykvcache{} in the bottom panel. The shaded area tracks CPU-pool occupancy, while the lines show disk read and write throughput. Native \vllm{} begins a promotion whenever a scheduler lookup finds KV data on disk. Several lookups can occur in quick succession, causing multiple promotions to reserve most of the CPU cache concurrently. Stores can then fail to obtain staging memory, and completed promotions may be evicted immediately to make room. When the request is eventually scheduled, its demand load may again find no free CPU memory and fall back to recomputing the prefix. The native trace repeatedly pins most of the CPU pool and continues beyond 1{,}200~s. In the traced run, only two requests ultimately transferred the promoted data from CPU to GPU, and a total of 3.4~TB of data was read from disk, while \pykvcache{} only read 85~GB, and the total cache size on each system was 465~GB. We reported this behaviour upstream as \href{https://github.com/vllm-project/vllm/issues/49902}{\vllm{} issue \#49902}. The issue has been acknowledged, and maintainers expect a policy or mechanism to detect back pressure, discussed in \href{https://github.com/vllm-project/vllm/issues/50031}{\#50031} and \href{https://github.com/vllm-project/vllm/pull/50014}{\#50014}.

\pykvcache{} instead permits only one speculative preload at a time and stops issuing preload work when a demand load arrives. It also bounds the memory reserved by each load or store to the configured I/O depth. In this configuration, \pykvcache{} needs only 576~MB of free CPU memory even when the complete context is much larger. The lower panel of \autoref{fig:pool-compare} shows that this bounded staging avoids the repeated occupancy spikes and allows the run to finish after approximately 480~s. This experiment shows that moving reads earlier is not sufficient by itself, and that speculative work should be treated for what it is: speculative, and it must also be bounded and yield to demand operations.

\subsection{Bailian Production Traces}

Finally, we replay the Coder, interactive (A), and API-driven (B) traces from Alibaba Cloud \bailian{}. As shown in \autoref{fig:bailian-ttft}, \pykvcache{} is 1.12--1.79$\times$ faster than GPU prefix caching and 1.10--1.25$\times$ faster than \lmcache{} on the local node (RTX~4000 Ada). It remains close to the native \vllm{} \kvoffloadname{} implementation, which is only 1.02--1.06$\times$ faster across the three traces. This is largely due to the smaller prompt sizes and lower prefix reuse of this workload, compared to \longbench{} and \scbench{}.

The same traces show a different outcome on the Snellius node. Their average request length is below the measured 6{,}203 token SSD break-even point for Qwen3~4B, on our setup, and the H100's larger GPU memory retains a large portion of the working set. As a result the TTFT of only using GPU prefix caching is identical to that of \pykvcache{}, while the native \vllm{} \kvoffloadname{} implementation is 0.2~s higher. Whether external caching is useful depends on both the workload's reusable-prefix distribution and the GPU's compute and memory capacity. It motivates \pykvcache{}'s model and hardware-specific break-even gate instead of loading every matched prefix.

\subsection{Evaluation Takeaways}

The evaluation shows that \pykvcache{} improves both the transfer path and the placement of disk reads. In the controlled disk-only benchmark, it halves TTFT relative to \lmcache{} at 80k tokens. Preload provides a further 1.34--1.66$\times$ improvement over the same engine without lookahead. \longbench{} and \scbench{} show that these gains extend to chained and multi-turn contexts, while the \scbench{} trace demonstrates why bounded staging and demand prioritization are required under memory pressure. The \bailian{} traces show the boundary of these benefits, where external KV caching helps on the smaller GPU, but on the H100 many prefixes are below break-even and should remain in GPU memory or be recomputed.

\section{Discussion}

The results suggest that external KV caching should be understood as a critical path optimization, rather than only as an extension of the memory hierarchy. Peak storage bandwidth determines how quickly KV data can be moved, but it does not determine when that movement begins, how much intermediate memory it reserves, or whether loading is faster than recomputation. Across the evaluated systems, these scheduling and resource management decisions were often as important as the underlying storage medium. This distinction explains why a slower source can occasionally produce a lower TTFT.

\subsection{Importance of the Critical Path}

The storage microbenchmarks reinforce this interpretation. The tested I/O engines differed substantially for 4~KiB requests, especially when additional threads were used, but converged near the device bandwidth limit for 1~MiB requests with one thread, as shown in \autoref{fig:io-engine-request-size}. The KV-cache workload generated mostly large requests, so increasing small-I/O operation rate or replacing the I/O engine did not by itself improve end-to-end TTFT. For this workload, the useful role of asynchronous I/O is therefore not to maximize IOPS. It is to maintain enough outstanding large operations to use the SSD bandwidth.

The GPU transfer results lead to a similar conclusion. \lmcache{} and the native \kvoffloadname{} implementation achieved comparable bulk GPU $\leftrightarrow$ CPU bandwidth, but the \offload{} path issued fewer, larger transfers, crucially hiding store work alongside model forward passes. Its advantage therefore came from transfer granularity and overlap rather than a substantially faster physical copy path. \pykvcache{} retains this \offload{} execution model and applies the same principle to storage reads. Moving a read earlier can reduce visible latency even when the number of transferred bytes is unchanged. In the controlled experiments, preload improved TTFT by 1.34--1.66$\times$ over the same \pykvcache{} engine without preload. This is evidence that optimizing when work executes can be more valuable than optimizing the isolated speed of that work.

Preloading converts demand work into speculative work. It is beneficial only when the selected request is likely to execute and remains queued long enough for useful I/O to complete. An unlimited preloader could consume storage bandwidth, staging memory, and CPU $\leftrightarrow$ GPU transfer capacity for requests that are delayed or never scheduled. The \scbench{} experiment demonstrates this risk, where several native promotions reserved most of the CPU tier concurrently, displaced recently promoted blocks, and left later demand operations without sufficient memory. \pykvcache{}'s policy of allowing one preload, sharing its buffers with demand transfers, and yielding when a demand load arrives is intentionally conservative. The result suggests that successful preloading requires admission control and resource bounds, not merely earlier submission. Unlike native \vllm{} \offload{}, which performs promotions based on lookups to KV blocks, \pykvcache{}'s access to the scheduler list allows it to perform smarter choices about which requests to preload.

\subsection{Comparison with Existing Systems}
\label{subsec:comparison}

\lmcache{} supports multiple serving engines, storage backends, and distributed deployments, whereas \pykvcache{} is specialized for \vllm{}'s \kvoffloadname{} and shared filesystem storage \cite{lmcache2025efficient,lmcacheArchitecture}. In the evaluated configuration, \pykvcache{} benefits from \offload{}'s coarser, deferred transfers but this does not make \offload{} universally superior or replace \lmcache{}'s broader functionality and extensive support for other storage media and distributed deployments. The native \vllm{} \offload{} implementation is the closest comparison because it uses the same API and integrated CPU and filesystem tiers \cite{vllmKVOffloading}. \pykvcache{} achieves similar performance to \vllm{}'s native \offload{} implementation in our tiered cache tests and performs better on \longbench{} and \scbench{}. The \scbench{} result reflects \pykvcache{}'s more conservative preloading policy, and its more robust design when under demanding worst-case conditions.

\pykvcache{} and \llmd{} both use hierarchical, hash-addressed files for sharing and atomic publication \cite{llmdFSKVCache}. \pykvcache{} uses transfer jobs bounded by I/O depth and a shared staging pool, instead of the separate buffers used by \llmd{}.

GPU prefix caching remains preferable when the working set fits in VRAM because it avoids any transfer or copy, since \vllm{} just swaps pointers.

\subsection{When External Caching Should Be Used}

The Pareto frontiers in \autoref{fig:other-kv-pareto} show why a cache hit alone is an insufficient policy signal. External reuse is most attractive when a request has a long reusable prefix, the reusable working set exceeds GPU capacity, and the request waits long enough for transfers to overlap with other work. It becomes less attractive for short prefixes, low reuse, faster GPUs, or lightly loaded systems with little queueing time. The appropriate threshold also changes with model architecture, KV datatype, GPU compute rate, PCIe bandwidth, storage bandwidth, and source tier. Consequently, the numerical thresholds measured in this work are properties of specific configurations.

The \bailian{} traces illustrate both sides of this operating envelope. On the local node, external caching improved TTFT because the smaller RTX~4000 Ada retained less of the working set and recomputation was comparatively expensive. On the H100 system, the larger VRAM capacity retained more prefixes and the average request was below the measured SSD break-even point for Qwen3~4B. External transfers consequently offered little benefit even when reusable data existed. A practical multi-tier policy should therefore use CPU or SSD only when their predicted transfer cost is lower than recomputation, and preserve recomputation as a valid fallback rather than treating it as a cache failure or miss.

Break-even-aware decisions should also include the expected placement of work. A static threshold based only on prefix length and device bandwidth cannot distinguish a demand read from a preload that may be hidden by queueing. Conversely, a preload that appears profitable from transfer time alone may be wasteful when the request is unlikely to run soon. A stronger policy would combine prefix size, source tier, current staging capacity, measured transfer rates of each medium, expected queue time. The bounded policy evaluated here is a first step toward a more complex scheduler and cache controller.

\section{Related Work}

\subsection*{GPU-resident Prefix Reuse}

\pagedattention{} makes non-contiguous KV blocks practical inside an LLM serving engine and underpins \vllm{}'s automatic prefix caching \cite{kwon2023efficient,vllmPrefixCaching}. SGLang's RadixAttention organizes reusable prefixes in a radix tree and integrates cache state with request scheduling \cite{zheng2024sglang}. These systems avoid external transfers when the reusable working set fits in GPU memory. \pykvcache{} addresses the complementary case in which reusable prefixes exceed that capacity.

\subsection*{External and Disaggregated KV Caches}

\lmcache{} provides reusable KV storage across serving engines and multiple local or remote backends \cite{lmcache2025efficient,lmcacheArchitecture}. MemServe separates KV storage from inference workers through an elastic memory pool and includes a cost model based on request and cache state \cite{hu2024memserve}. Mooncake similarly treats KV transfer as a central part of disaggregated serving and coordinates prefill, decode, and cache resources around a KV-centric data path \cite{qin2025mooncake}. DualPath extends this line by observing that the storage NICs attached to prefill engines saturate while those attached to decode engines stay idle, and adds a second load path that pulls KV state into decode engines and forwards it to prefill over the compute network, reporting up to 1.87$\times$ higher offline inference throughput on agentic workloads \cite{wu2026dualpath}. These systems target broader distributed deployments than \pykvcache{}.

\subsection*{SSD KV Caching}

IMPRESS states the premise this work measures: when prefix KV state must be stored on disk, reusing it does not always reduce TTFT, because disk latency can exceed the prefill it saves \cite{chen2025impress}. It answers by loading less, restoring only the tokens it judges important from the similarity of their index sets across attention heads. \pykvcache{} answers the same observation by deciding whether to load at all rather than which parts to load. The two are compatible: selective loading moves fewer bytes and therefore shifts the boundary, but something must still decide when even the reduced transfer is not worth issuing.

Bidaw is the closest published relative of the break-even gate. It targets the same two-tier arrangement of host memory and SSDs evaluated here, and weighs storage footprint against computational saving when deciding what to retain \cite{hu2026bidaw}. Bidaw selects what to keep, while \pykvcache{} decides whether a load already matched in the cache is worth placing on the critical path against the cost of recomputing the same prefix. Tutti is concurrent work on the same tier, eliminating CPU intervention from the I/O control path through a GPU-centric object store and slack-aware scheduling \cite{qiu2026tutti}. Tutti is also built on \vllm{}, so the difference from \pykvcache{} is the I/O path rather than the serving engine: \pykvcache{} keeps the CPU-staged path and asks instead whether the transfer is worth issuing.

Several systems converge on transfer granularity as the dominant SSD-side concern, whether by consolidating KV pairs into larger blocks, adopting a single granularity for pruning and prefetching alike, or spreading co-activated entries across multiple devices \cite{zheng2026solidattention,zou2026contiguouskv,wang2026swarm}. KV compression helps by increasing effective tier capacity, and shifting the measured break-even frontier \cite{liu2024kivi,liu2024cachegen}. Where memory pressure is severe enough to make page-cache behaviour unpredictable, DUAL-BLADE abandons the filesystem altogether for an NVMe-direct path \cite{jeong2026dualblade}.

\subsection*{Prefetching and Lookahead}

Prefetching recurs as the mechanism for moving KV transfers off the critical path before a request is scheduled. CachedAttention performs layer-wise preloading and asynchronous saving with scheduler-aware fetching and eviction \cite{gao2024cachedattention}. Other systems also perform prefetching, by looking at the scheduler queue \cite{wang2026pcr} or by speculating required KV entries \cite{lee2024infinigen}. HyMCache performs a similar action, but uses a CXL hybrid memory device to stage KV in device-side DRAM ahead of the foreground read \cite{jang2026hymcache}. \pykvcache{} takes the signal directly from the scheduler's waiting queue, so nothing is predicted.
\section{Limitations and Future Work}

\subsection{Limitations}

The experiments cover only two hardware classes, a small number of models, and a limited set of SSD and storage configurations. Notably, they exclude networked and RAID configurations. Our runs only use FP16 KV data and one output token in order to isolate prefill and cache movement. Although varying output length in the mixed workload did not appear to change much in our tests, the study does not establish effects on sustained decode throughput or multi GPU configurations. All experiments use a single KV block size of 256 tokens, so the results do not show how staging, file sizes and transfer granularity behave at other block sizes.

The trace workloads broaden the evaluation beyond fixed synthetic prefixes, but replay does not reproduce every aspect of a live production deployment.

Every measurement in this work uses a CPU dependent I/O path. Data is staged in host memory when transferring in/out of the GPU. Some research has explored direct GPU access to CPU memory or storage for both KV and the forward pass, and can dramatically change our conclusions \cite{luo2026directkv,qureshi2023bam}. A faster path does not remove the break-even between loading and recomputing, but it does move it, so the frontier we measure belongs to the path we evaluate.

Finally, \lmcache{} and \vllm{} are rapidly evolving systems. The results describe the versions and configurations evaluated here. The comparative conclusions should therefore be read as evidence about design of such systems rather than as permanent rankings of projects.

\subsection{Future Work}

A dynamic serving engine could continuously estimate prefill time, tier bandwidth, queue delay, and promotion success, then update store, load, and defer decisions as the workload changes. Preload selection could similarly move beyond a single request, to a small priority queue governed by explicit memory and I/O budgets. The lookahead signal could also be pushed below the serving engine. A prefix-cache lookup already yields the ordered list of blocks that a request will consume, and recent work shows that a memory device can consume that ordering directly to stage data ahead of the foreground read \cite{jang2026hymcache}. The same signal could drive readahead in the storage layer rather than in the connector. Other useful directions include combining external caching with KV quantization or compression, testing remote and distributed filesystems, and evaluating multi node, multi GPU and small cluster configurations to test cache sharing under failures and contention. A broader evaluation should include tail TTFT, throughput, fairness, energy, and cost. Together, these extensions would turn the central result of this work into a general policy determining when and how KV data moves or whether it should move at all.

\section{Conclusion}

External KV caching is useful only when a reusable prefix can be moved sooner than it can be recomputed. Our characterization shows that this comparison depends not only on storage bandwidth but also on overlap between compute and copy operations. These findings led to \pykvcache{}, a \vllm{} \kvoffloadname{} connector that uses asynchronous direct I/O, bounded shared staging, scheduler-aware preload, and setup-specific break-even decisions.

In the evaluated configurations, \pykvcache{} halves disk-only query TTFT relative to \lmcache{} at 80k tokens and remains within 4\% of native \vllm{} in the multi-tier comparison. \longbench{} and \scbench{} show benefits when reusable contexts exceed GPU capacity, while the \bailian{} traces on the H100 show the opposite boundary, where short prefixes and a large GPU leave little reason to use an external tier. External caching pays off when admission selects useful reuse and scheduling removes transfer work from the request's critical path.

\section*{Acknowledgment}
We thank Zebin Ren and the AtLarge group at VU Amsterdam for their support.

{\footnotesize \bibliographystyle{acm}
\bibliography{refs}}

\textbf{Notes:} IBM is a trademark of International Business Machines Corporation, registered in many jurisdictions worldwide. Intel and Intel Xeon are trademarks or registered trademarks of Intel Corporation or its subsidiaries in the United States and other countries. Linux is a registered trademark of Linus Torvalds in the United States, other countries, or both. Java and all Java-based trademarks and logos are trademarks or registered trademarks of Oracle and/or its affiliates. Other products and service names might be trademarks of IBM or other companies.

\end{document}